\documentclass[twocolumn]{aastex631}

\usepackage{xspace}
\usepackage{natbib}
\usepackage{enumitem}

\newcommand{\halpha}{\ensuremath{\textrm{H}\alpha}\xspace}
\newcommand{\hbeta}{\ensuremath{\textrm{H}\beta}\xspace}
\newcommand{\OIII}{\ensuremath{[\mathrm{O}\textsc{ iii}]}\xspace}
\newcommand{\NII}{\ensuremath{[\mathrm{N}\textsc{ ii}]}\xspace}
\newcommand{\OII}{\ensuremath{[\mathrm{O}\textsc{ ii}]}\xspace}
\newcommand{\HII}{\ensuremath{\mathrm{H}\textsc{ ii}}\xspace}

\newcommand{\AV}{\ensuremath{\mathrm{A}_{\mathrm{V}}}\xspace}
\newcommand{\ABalmer}{\ensuremath{\mathrm{A}_{\mathrm{Balmer}}}\xspace}

\newcommand{\Msun}{\ensuremath{M_{\odot}}\xspace}
\newcommand{\Mstar}{\ensuremath{M_{*}}\xspace}
\newcommand{\logoh}{\ensuremath{[12 + \log_{10}\left(\mathrm{O/H}\right)]}\xspace}

\shorttitle{Dust Attenuation in $1.3\leq z\leq 2.6$ Star-forming Galaxies from MOSDEF}
\shortauthors{Lorenz et al.}
\graphicspath{./}
\begin{document}

\title{An Updated Dust-to-Star Geometry: Dust Attenuation Does Not Depend on Inclination in $1.3\leq z\leq 2.6$ Star-Forming Galaxies from MOSDEF}

\author[0000-0002-5337-5856]{Brian Lorenz}
\affiliation{Department of Astronomy, University of California, Berkeley, CA 94720, USA}
\author[0000-0002-7613-9872]{Mariska Kriek}
\affiliation{Leiden Observatory, Leiden University, P.O. Box 9513, 2300 RA Leiden, The Netherlands}
\author[0000-0003-3509-4855]{Alice E. Shapley}
\affiliation{Department of Physics and Astronomy, University of California, Los Angeles, 430 Portola Plaza, Los Angeles, CA 90095, USA}
\author[0000-0001-9687-4973]{Naveen A. Reddy}
\affiliation{Department of Physics and Astronomy, University of California, Riverside, 900 University Ave., Riverside, CA 92521, USA}
\author[0000-0003-4792-9119]{Ryan L. Sanders}
\affiliation{Department of Physics and Astronomy, University of California, Davis, One Shields Ave, Davis, CA 95616, USA}
\author[0000-0001-6813-875X]{Guillermo Barro}
\affiliation{University of the Pacific, Stockton, CA 90340, USA}
\author[0000-0002-2583-5894]{Alison L. Coil}
\affiliation{Center for Astrophysics and Space Sciences, Department of
Physics, University of California, San Diego, 9500 Gilman Dr., La
Jolla, CA 92093, USA}
\author[0000-0001-5846-4404]{Bahram Mobasher}
\affiliation{Department of Physics and Astronomy, University of California, Riverside, 900 University Ave., Riverside, CA 92521, USA}
\author[0000-0002-0108-4176]{Sedona H. Price}
\affiliation{Max-Planck-Institut für extraterrestrische Physik (MPE), Giessenbachstr. 1, D-85748 Garching, Germany}
\author[0000-0003-4852-8958]{Jordan N. Runco}
\affiliation{Department of Physics and Astronomy, University of California, Los Angeles, 430 Portola Plaza, Los Angeles, CA 90095, USA}
\author[0000-0003-4702-7561]{Irene Shivaei}
\affiliation{Steward Observatory, University of Arizona, Tucson, AZ 85721, USA}
\author[0000-0002-4935-9511]{Brian Siana}
\affiliation{Department of Physics and Astronomy, University of California, Riverside, 900 University Ave., Riverside, CA 92521, USA}
\author[0000-0002-6442-6030]{Daniel R. Weisz}
\affiliation{Department of Astronomy, University of California, Berkeley, CA 94720, USA}

%% Note that the \and command from previous versions of AASTeX is now
%% depreciated in this version as it is no longer necessary. AASTeX 
%% automatically takes care of all commas and "and"s between authors names.

%% AASTeX 6.31 has the new \collaboration and \nocollaboration commands to
%% provide the collaboration status of a group of authors. These commands 
%% can be used either before or after the list of corresponding authors. The
%% argument for \collaboration is the collaboration identifier. Authors are
%% encouraged to surround collaboration identifiers with ()s. The 
%% \nocollaboration command takes no argument and exists to indicate that
%% the nearby authors are not part of surrounding collaborations.

%% Mark off the abstract in the ``abstract'' environment. 
\begin{abstract}

We investigate dust attenuation and its dependence on viewing angle for 308 star-forming galaxies at $1.3\leq z\leq2.6$ from the MOSFIRE Deep Evolution Field (MOSDEF) survey. We divide galaxies with a detected \halpha emission line and coverage of \hbeta into eight groups by stellar mass, star formation rate (SFR), and inclination (i.e., axis ratio), then stack their spectra. From each stack, we measure Balmer decrement and gas-phase metallicity, then we compute median \AV and UV continuum spectral slope ($\beta$). First, we find that none of the dust properties (Balmer decrement, \AV, $\beta$) vary with axis ratio. Second, both stellar and nebular attenuation increase with increasing galaxy mass, showing little residual dependence on SFR or metallicity. Third, nebular emission is more attenuated than stellar emission, and this difference grows even larger at higher galaxy masses and SFRs. Based on these results, we propose a three-component dust model where attenuation predominantly occurs in star-forming regions and large, dusty star-forming clumps, with minimal attenuation in the diffuse ISM. In this model, nebular attenuation primarily originates in clumps, while stellar attenuation is dominated by star-forming regions. Clumps become larger and more common with increasing galaxy mass, creating the above mass trends. Finally, we argue that a fixed metal yield naturally leads to mass regulating dust attenuation. Infall of low-metallicity gas increases SFR and lowers metallicity, but leaves dust column density mostly unchanged. We quantify this idea using the Kennicutt-Schmidt and fundamental metallicity relations, showing that galaxy mass is indeed the primary driver of dust attenuation.
\end{abstract}

%% Keywords should appear after the \end{abstract} command. 
%% The AAS Journals now uses Unified Astronomy Thesaurus concepts:
%% https://astrothesaurus.org
%% You will be asked to selected these concepts during the submission process
%% but this old "keyword" functionality is maintained in case authors want
%% to include these concepts in their preprints.

%% From the front matter, we move on to the body of the paper.
%% Sections are demarcated by \section and \subsection, respectively.
%% Observe the use of the LaTeX \label
%% command after the \subsection to give a symbolic KEY to the
%% subsection for cross-referencing in a \ref command.
%% You can use LaTeX's \ref and \label commands to keep track of
%% cross-references to sections, equations, tables, and figures.
%% That way, if you change the order of any elements, LaTeX will
%% automatically renumber them.
%%
%% We recommend that authors also use the natbib \citep
%% and \citet commands to identify citations.  The citations are
%% tied to the reference list via symbolic KEYs. The KEY corresponds
%% to the KEY in the \bibitem in the reference list below. 

\section{Introduction} \label{sec:intro}

Understanding the effects of dust attenuation on galaxy observations is crucial for accurately determining their physical parameters. Measurements of fundamental parameters such as mass and star formation rate (SFR) may vary significantly depending on the assumed dust properties and distribution \citep[][and references therein]{salim_dust_2020}. Consequently, it is essential to have a robust understanding of how the effects of dust correlate with various galaxy properties.

A popular model for the for the dust-to-star geometry describes it with two components: dust in the interstellar medium (ISM) that attenuates all light, and more optically thick dust within star-forming regions that primarily affects young stars \citep[e.g.,][]{charlot_simple_2000}. Dust in star-forming regions can be observed through the Balmer decrement, the ratio of \halpha emission line flux to \hbeta. On the other hand, the effects of ISM dust is most apparent from its effect on the shape of the stellar continuum. This two-component model is supported by observations. In particular, \citet{calzetti_dust_2000} found that the dust in nearby starburst galaxies is a combination of a ``warm'' component (originating from the star-forming regions) and ``cool'' component (originating from the ISM) with different optical depths. 

However, at higher redshift, there are conflicting results on the effectiveness of the two-component model. For star-forming galaxies at $z\approx 1.4$, \citet{price_direct_2014} find extra dust toward \HII regions compared to the stellar attenuation, in agreement with a two-component dust model. Furthermore, the authors find that the difference between the nebular and stellar attenuation decreases with increasing sSFR (SFR/M$_*$). This result is an additional prediction of the two-component model; as sSFR goes down, the stellar light is increasingly dominated by stars attenuated only by the diffuse ISM, and thus the observed stellar attenuation decreases relative to the nebular attenuation. In contrast, \citet{reddy_mosdef_2015} show that for star-forming galaxies at $1.4\leq z\leq 2.6$, the difference between the nebular and stellar attenuation increases with SFR. They propose that galaxies with low SFRs have nebular lines only attenuated by the dust in the diffuse ISM, while galaxies with high SFRs have nebular line attenuation dominated by highly obscuring dust around star-forming regions. However, this additional dust is patchy and only affects some star-forming regions, so the diffuse ISM component still dominates the stellar attenuation. Consequently, the geometry proposed by \citet{reddy_mosdef_2015} leads to a large difference in Balmer and stellar attenuation. Hence, these two works at similar redshifts propose different dust-to-star geometry models. Other dust-to-star geometry models have also been proposed. For example, \citet{reddy_mosdef_2020} proposes that the youngest, brightest stars are most heavily obscured and \citet{fetherolf_mosdef_2023} proposes a model in which galaxy mass affects the dust mixing timescale, and therefore the attenuation.

Galaxy inclination is a powerful tool to further assess the dust-to-star geometry at $z\approx 2$. For a simple two-component model \citep[e.g.,][]{charlot_simple_2000, price_direct_2014}, we expect that the stellar continuum attenuation might vary with the inclination of a disky galaxy, since stellar light should go through more dust in a system that is viewed edge-on. On the other hand, if the obscuration is dominated by dust in the star-forming regions, then the path length does not depend on inclination. In that case, the Balmer decrement would not vary as strongly with inclination. However, for a single dust-component model, or for the low SFR galaxies in the model by \citet{reddy_mosdef_2015}, we would expect Balmer decrement to increase for more inclined galaxies due to attenuation by ISM dust.

At low redshift, galaxy inclination has proved to be an interesting lens to study dust properties. In a sample of SDSS star-forming disk galaxies at $z\approx 0.07$, \citet{yip_extinction_2010} found that Balmer decrement does not depend on inclination, while the stellar continuum attenuation increases in edge-on galaxies. Additionally, they find that the Balmer-line color excess is larger than the stellar continuum color excess at all inclinations. Both of these results support a two-component model, where the dust affecting the star-forming regions is distributed differently than dust affecting the stellar continuum. This effect is also described in \citet{wild_empirical_2011}. Along the same lines, \citet{battisti_characterizing_2017} observes no correlation between Balmer optical depth and inclination for a sample of GALEX star-forming galaxies at $z<0.105$, but they find a clear correlation between the dust attenuation curve and inclination.

Studying the relationship between dust properties and galaxy inclination at $z\approx 2$ to shed light on the observed inconsistencies requires a sample of distant galaxies with measured inclinations and prominent emission lines from which we can derive dust properties. Such analysis also requires accurate measurements of galaxy parameters such as stellar mass and SFR. Fortunately, by combining the deep spectra from the MOSFIRE Deep Evolution Field survey \citep{kriek_mosfire_2015} with photometry and imaging from CANDELS \citep{grogin_candels_2011, koekemoer_candels_2011}, we now have a sample of high-redshift galaxies for which these detailed dust measurements are possible. 

In this work, we use the MOSDEF survey to investigate how dust properties are affected by galaxy inclination, which will improve our understanding of dust-to-star geometry at $z\approx 2$. In Section \ref{sec:data}, we describe the MOSDEF survey and our selection of a sub-sample for the analysis of dust properties and inclination. In Section \ref{sec:data_analysis}, we detail our analysis techniques, including spectral stacking and emission line measurements. In Section \ref{sec:results}, we present our findings on the relationship between the measured dust properties (Balmer decrement, \AV, $\beta$) and galaxy axis ratio. Finally, in Section \ref{sec:discussion}, we propose a three-component dust model that explains our results. 

Throughout this work we assume a $\Lambda$CDM cosmology with $\Omega_m=0.3$, $\Omega_{\Lambda}=0.7$, and $H_0=70$ km s$^{-1}$ Mpc$^{-1}$.

\section{Data} \label{sec:data}

\subsection{The MOSDEF Survey} \label{subsec:mosdef_survey}

For this work, we use galaxy spectra from the MOSFIRE Deep Evolution Field (MOSDEF) Survey \citep{kriek_mosfire_2015}. MOSDEF obtained moderate resolution ($R=3000-3650$) rest-frame optical spectra of $\sim$$1500$ H-band selected galaxies in the CANDELS fields \citep{grogin_candels_2011, koekemoer_candels_2011}. Observations were taken in the $Y, J, H,$ and $K$ bands using the MOSFIRE spectrograph \citep{mclean_mosfire_2012} on the Keck I telescope from December 2012 to May 2016. The sample spans three redshift intervals ($1.37\leq z\leq 1.70, 2.09\leq z\leq 2.61,$ and $2.95\leq z \leq 3.60$), targeted such that many prominent rest-frame emission and absorption features fall into the observed atmospheric transmission windows.  For complete details of the MOSDEF survey, including sample selection, observation details, data reduction, and initial properties, we refer to \citet{kriek_mosfire_2015}.

We aim to study the effects of galaxy inclination on dust parameters, so we need robust axis ratio measurements. Structural parameters for all galaxies in MOSDEF, including axis ratio ($b$/$a$), were measured by \citet{van_der_wel_structural_2012} using GALFIT \citep{peng_detailed_2010}. Fits were made to HST imaging in three bands (F125W, F140W, and F160W) from CANDELS (dataset available via \dataset[DOI: 10.17909/T94S3X]{10.17909/T94S3X}). 

In addition to spectra from MOSDEF, we have access to deep photometry from the 3D-HST survey \citep{brammer_3d-hst_2012, skelton_3d-hst_2014, momcheva_3d-hst_2016}. Stellar masses are determined by fitting the emission-line corrected multi-band photometry ($0.3-8.0 \mu m$) using the FAST fitting code \citep{kriek_ultra-deep_2009}. We adopt flexible stellar population models \citep[FSPS,][]{conroy_propagation_2009, conroy_propagation_2010}, MOSFIRE spectroscopic redshifts, a Chabrier stellar initial mass function (IMF) \citep{chabrier_galactic_2003}, the \citet{calzetti_dust_2000} attenuation curve, delayed exponentially declining star formation histories, and solar metallicity. These best-fit models also yield measurements of dust attenuation in the V-band (\AV) and UV slope ($\beta$).

Prominent emission lines (\halpha, \hbeta, \OIII, \NII, \OII) from MOSDEF spectra were fit to determine individual galaxy properties such as SFR and metallicity. Most emission lines were measured by fitting Gaussian and linear components \citep{reddy_mosdef_2015, kriek_mosfire_2015}. For Balmer lines, stellar absorption must be modeled to ensure accurate flux measurements, so the best-fit stellar population models are used to measure the Balmer absorption lines and correct the Balmer emission. For full details on emission line fits to the individual galaxies, we refer to \citet{reddy_mosdef_2015}.

\begin{figure*}[tp]
\vglue -5pt
\centering
\includegraphics[width=0.95\textwidth]{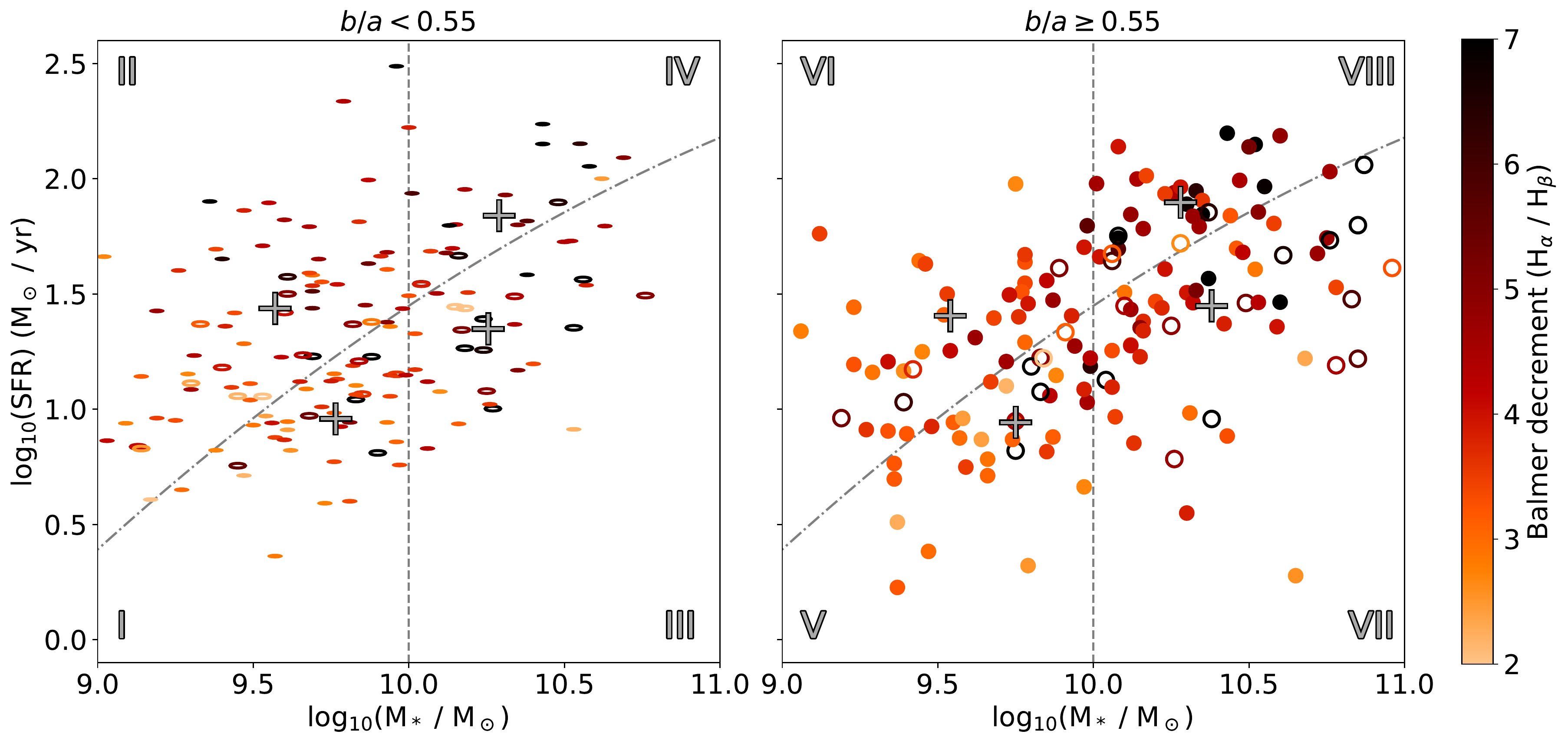}
\caption{
SFR vs. stellar mass of star-forming galaxies in our sub-sample from the MOSDEF survey at $1.37\leq z\leq 2.61$, split into edge-on galaxies ($b/a<0.55$) (left, ellipses) and face-on galaxies ($b/a\geq 0.55$) (right, circles). These symbols will be used throughout this work to denote edge-on (ellipses) and face-on (circles). The dashed lines show the divisions into groups by mass and the \citet{whitaker_constraining_2014} fit to the star-forming main sequence scaled to our sample. Symbols are color-coded by their initial measurement of Balmer decrement, with galaxies that have lower-limits on their Balmer decrement displayed as open points. The median stellar mass and SFR for each of the eight groups is denoted with a grey cross. From this figure, it is already apparent that higher mass galaxies have higher Balmer decrements.
}
\label{fig:sample_cut}
\end{figure*}

\subsection{SFR Measurement}

For our analysis, it is important to have an accurate SFR measurement for all galaxies in our sample. All galaxies in MOSDEF that have a strong $\left(3\sigma\right)$ detection in \halpha and \hbeta have a measured dust-corrected SFR \citep{reddy_mosdef_2015, shivaei_mosdef_2015}. In summary, the measured \halpha flux is dust-corrected using the Balmer decrement (\halpha/\hbeta) and the \cite{cardelli_relationship_1989} extinction curve \cite[shown to work well for MOSDEF galaxies by][]{reddy_mosdef_2020}, then converted to \halpha luminosity using the measured redshift. SFRs are then determined from \halpha luminosity assuming a \citet{chabrier_galactic_2003} IMF and the relation from \citet{hao_dust-corrected_2011} for solar metallicity. In this work, we adopt these measured SFRs. However, we only include galaxies with at least a $5\sigma$ \halpha detection for a more robust analysis and measurement of emission line properties after stacking (see Section \ref{subsec:sample_select}).

For the galaxies that have a $5\sigma$ \halpha measurement but no detected \hbeta emission, we can only estimate a lower limit on the Balmer decrement using the previous method, and therefore a lower limit on the SFR. Instead, for these galaxies that lack significant \hbeta detections, we estimate SFR based on \halpha alone, correcting for dust using the \AV from the best-fit FAST models. We use a \citet{calzetti_dust_2000} extinction law to correct from the V band at $5500$\AA\ to the wavelength of \halpha at $6565$\AA. Finally, using the SFR calculation from \citet{hao_dust-corrected_2011} and assuming a \citet{chabrier_galactic_2003} IMF, we compute the SFR. We performed this calculation for all galaxies in the sample, including the galaxies that had strong \hbeta detections. The SFRs measured from both methods agreed well, with an offset of 0.1 dex and similar scatter \citep[see also][]{price_direct_2014}.

\subsection{Sample Selection} \label{subsec:sample_select}

For this work, we need a complete sample of star-forming galaxies with measured masses, SFRs, and axis ratios. To carry out measurements of emission features, we require galaxies in our sample to have a detected \halpha line and a covered \hbeta line. Consequently, we restrict our sample to $1.37\leq z\leq 2.61$ so that the \halpha and \hbeta emission features are targeted for all galaxies. This reduces the original $\sim$1500 galaxies to 1061. 

We then apply the following selection criteria to limit the sample to star-forming galaxies with accurate redshifts, SFRs, and little contamination from other sources. First, we remove any galaxies that do not have a high quality redshift measurement from MOSDEF, as we need to be able to accurately measure the location of emission features (114 galaxies). Some galaxies do not have coverage of the H$\alpha$ (43) or H$\beta$ (82) emission lines, which are essential for our measurement of Balmer decrements. Similarly, galaxies that do not have at least a signal-to-noise ratio of 5 for their \halpha measurement are discarded (267). In addition, any galaxies classified as AGN in MOSDEF are removed \citep{coil_mosdef_2015, azadi_mosdef_2017}, ensuring that any emission lines are almost entirely associated with star formation (144). Next, we remove galaxies that have their axis ratio measurements flagged as suspicious, bad, or non-existent by GALFIT and have differences between F125W and F160W axis ratios greater than 0.08 (15). We restrict our sample to galaxies with $9\leq \log_{10}(\Mstar/\Msun) \leq 11$, since the sample is only complete down to a stellar mass of $10^9\Msun$ (25). Finally, any galaxy where \halpha or \hbeta is contaminated by a sky line is removed, where sky lines are identified by pixels where the measured error is at least 10 times larger than the median measured error of the spectrum (63). After this process, we have 308 galaxies remaining in our sample, with masses, SFRs, and Balmer decrements or lower limits displayed in Figure \ref{fig:sample_cut}. All remaining galaxies fall in the star-forming region of the UVJ diagram (see Section \ref{sec:results}) \citep[e.g.,][]{wuyts_what_2007, williams_detection_2009}.

We note that these selection criteria may introduce some biases into our sample. In particular, we caution that the sample may be missing the most heavily obscured systems, since a signal-to-noise ratio of 5 is required for their \halpha measurement.

\begin{figure}[t]
\centering
\includegraphics[width=0.49\textwidth]{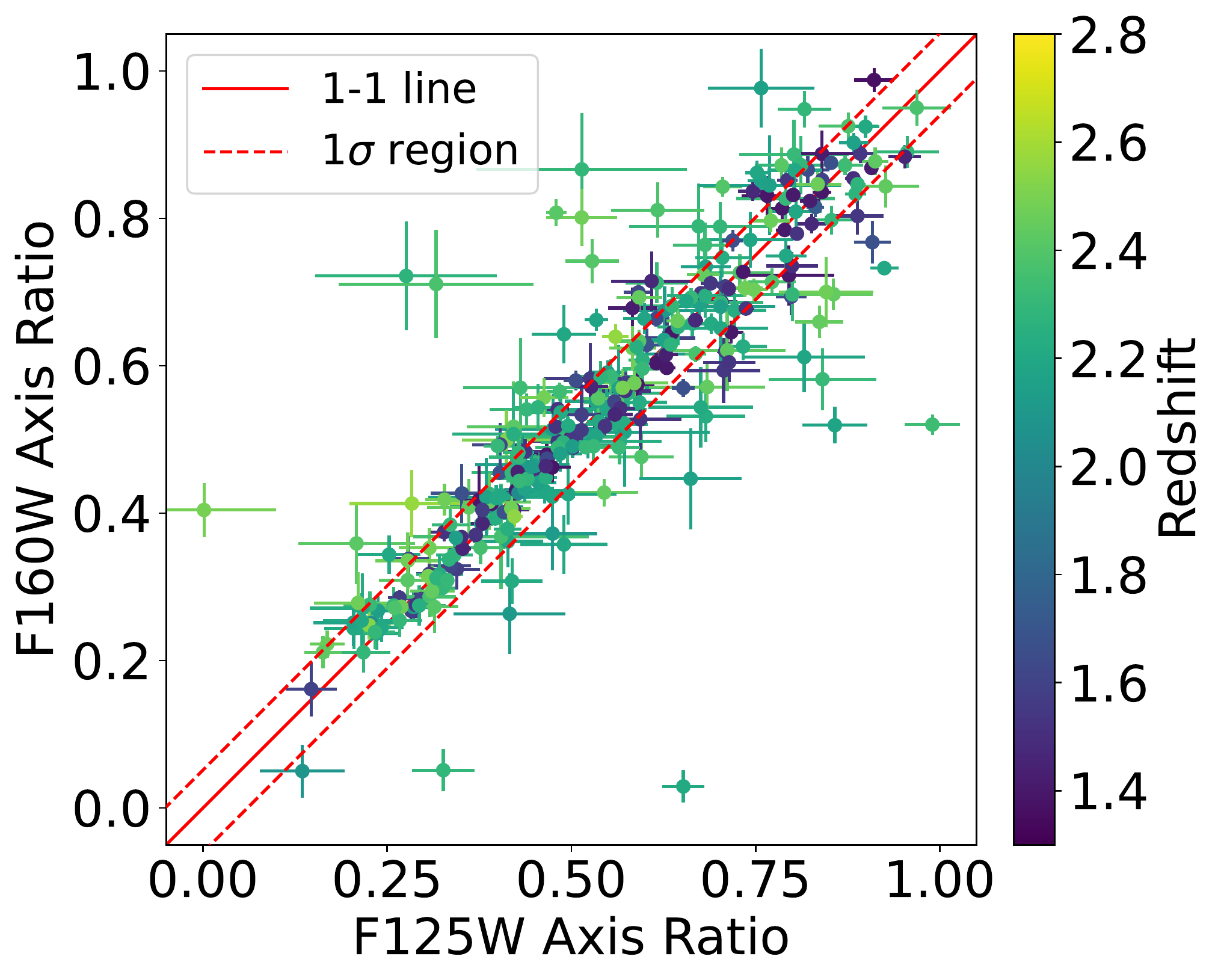}
\caption{
Comparison between measured F125W and F160W axis ratios for the galaxies in our sample, color-coded by redshift. A one-to-one line is drawn in solid red, and dashed lines show the 16th and 84th percentiles of the distribution of deviations from the one-to-one line. The measurements are well-correlated, so there is very little ambiguity between edge-on and face-on when dividing into two axis-ratio groups. We choose to use F160W axis ratios for galaxies in the higher redshift group ($z>2.09$) and F125W axis ratios for galaxies in the lower redshift group ($z<1.7$) so that galaxies are classified at similar rest-frame wavelengths. 
}
\label{fig:axis_ratio_compare}
\end{figure}

\begin{figure}[h]
\centering
\includegraphics[width=0.49\textwidth]{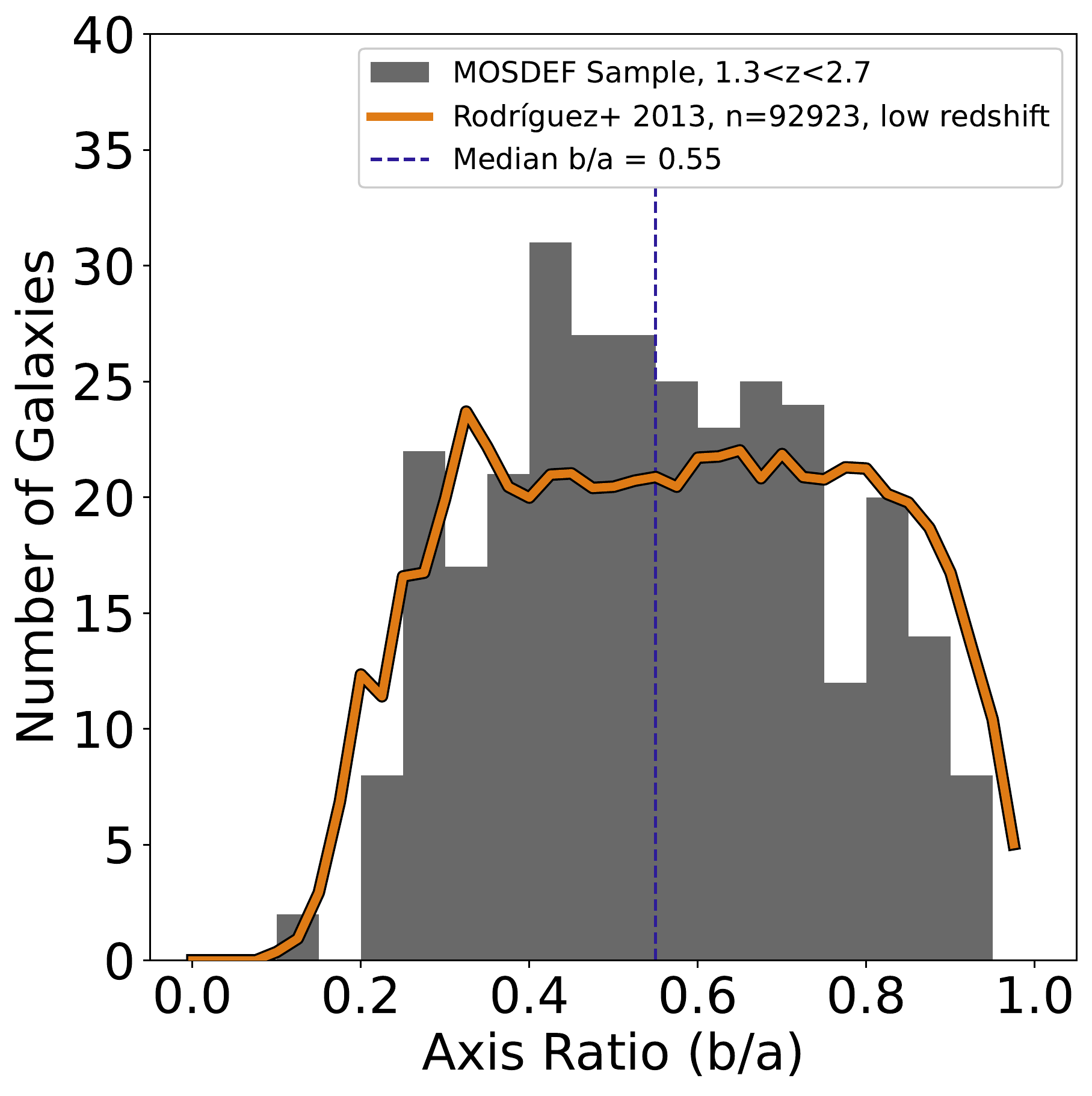}
\caption{
Histogram of the axis ratios for all galaxies in our sample, with a low-redshift, star-forming, disky comparison sample from \citet{rodriguez_intrinsic_2013} in orange. The blue dashed line shows the division between edge-on and face-on galaxies by the median axis ratio $\left(b/a=0.55\right)$. The observed distribution is consistent with the expected relatively flat distribution for disk galaxies, and inconsistent with the expected rising distribution for elliptical galaxies. 
}
\label{fig:ar_histogram_use_ratio}
\end{figure}

\subsection{Axis Ratios}
\label{subsec:axis_ratios}

To determine the effects of inclination on dust attenuation, we require accurate axis ratio measurements for all galaxies. As mentioned in Section \ref{subsec:mosdef_survey}, axis ratios $\left(b/a\right)$ were measured in F125W, F140W, and F160W by \citet{van_der_wel_structural_2012} using GALFIT \citep{peng_detailed_2010}. Due to their higher coverage and signal-to-noise, we limit our analysis to F125W and F160W axis ratio measurements. We checked the consistency of F125W and F160W axis ratios, and, for the vast majority of galaxies, they are consistent to better than 0.1 (see Figure \ref{fig:axis_ratio_compare}). Therefore, either F125W or F160W axis ratios should be a robust way to split the galaxies into groups as described in Section \ref{subsec:group_formation}. Since the galaxies come from two distinct redshift samples, $1.37\leq z\leq 1.70$ and $2.09\leq z\leq 2.61$, using only one filter would measure axis ratios at a wide range of rest-frame wavelengths. To better compare the galaxies, we use F125W axis ratios for the lower redshift sample (resulting in a median rest-wavelength of 4900\AA), and F160W axis ratios for the higher redshift sample (resulting in a median rest-wavelength of 4700\AA).

\subsection{Galaxy Morphology}
\label{subsec:morphology}

For our work, we examine galaxy parameters as a function of galaxy inclination. We use axis ratio as a proxy of galaxy inclination, which assumes that all galaxies are disks. Therefore, we must ensure that galaxies in our sample are consistent with being disky. Given the axis ratio distribution of galaxies, we can infer properties of the population. As described in \citet{lambas_true_1992}, a randomly oriented distribution of disk galaxies is expected to have a relatively flat distribution of axis ratios, while a randomly oriented distribution of ellipticals will have an increasing distribution that peaks at roughly $(b/a) =0.8$. Figure \ref{fig:ar_histogram_use_ratio} shows the distribution of axis ratios for the galaxies in our sample, which is consistent with the expected distribution for disk galaxies and inconsistent with the expected distribution for elliptical galaxies. Finally, \citet{price_mosdef_2016, price_mosdef_2020} showed that the star-forming galaxies in MOSDEF at $z\approx 1.5$ and $z\approx 2.2$ have an average $v/\sigma$ of $\approx3.5$ and $\approx2.0$ respectively, consistent with being rotating disks. Thus, we can assume in this work that galaxies are disky, so axis ratio is a good indicator of a galaxy's inclination. 

We also note that at this epoch of peak star formation, galaxy disks are thick \citep{elmegreen_observations_2006} and turbulent \citep{genzel_rings_2008, simons_z2_2017}, unlike the more rotationally supported disks that are observed locally. We further explore the implications of disk morphology as we discuss our results (Section \ref{subsec:implications_dust}).

\section{Data Analysis} \label{sec:data_analysis}

Our goal in this work is to measure the effects of galaxy inclination on dust attenuation for a complete sample of distant star-forming galaxies. However, many galaxies do not have a detected \hbeta line in their spectra, which is necessary to measure a Balmer decrement. By stacking spectra from galaxies with similar properties, we can dramatically increase the signal-to-noise ratio of the spectra and measure emission features, such as \hbeta, which were undetected in individual spectra.

\subsection{Forming Groups} \label{subsec:group_formation}

Since we require accurate emission line measurements to determine Balmer decrements, we divide the galaxies into eight groups with similar properties and stack their spectra. Many galaxy properties are known to vary strongly with mass and SFR, such as metallicity, morphology, and kinematics. Therefore, we ensure that galaxies of similar masses and SFRs are stacked together. For this sample, we divide the SFR-M$_*$ plane into 4 bins, as shown in Figure \ref{fig:sample_cut}. In particular, the galaxies are divided by mass into $9\leq \log_{10}(\Mstar/\Msun) < 10$ (which we will now refer to as ``low-mass'') and $10\leq \log_{10}(\Mstar/\Msun) < 11$ (referred to as ``high-mass''). Then, to finish dividing the SFR-M$_*$ plane, we use the star-forming main sequence (SFMS) rather than SFR alone. The SFMS is a tight relation between the SFR and mass of galaxies that varies with redshift. A galaxy's position compared to the SFMS is an indicator of how quickly it is forming stars relative to galaxies of similar mass and redshift. We linearly scale the SFMS fit from \citet{whitaker_constraining_2014} to our data, which results in: 
\begin{equation}\label{eq:sfms}
    \log_{10}\textup{SFR}\left(\frac{\Msun}{\textup{yr}}\right) = a + b\log_{10}\left(\frac{\Mstar}{\Msun}\right) - c\log_{10}\left(\frac{\Mstar}{\Msun}\right)^2
\end{equation}
where $a=-24.0415$, $b=4.1693$, and $c=0.1638$.
To ensure that there was no difference in the SFMS between the edge-on and face-on groups of galaxies, we fit the SFMS again to each of the groups separately, and found that the separate fits have very similar values to those of the total sample. Additionally, we recover the known redshift dependence of the SFMS by fitting the low redshift and high redshift samples separately. We find similar results whether we use a redshift-dependent SFMS or an SFMS that is constant with redshift. Therefore, for simplicity, we use the constant SFMS.  We divide the sample into two groups of SFR using Equation \ref{eq:sfms}, with one group above and one group below the SFMS. 

Finally, since our motivating question is how galaxy inclination affects dust properties, we further divide each of the 4 bins in the SFR-M$_*$ plane into two equal-sized groups in axis ratio, using the median axis ratio of the sample, $b/a = 0.55$, as a divider. Given that the galaxies are consistent with being disks (see Section \ref{subsec:morphology}), these two samples roughly correspond to ``edge-on'' (which will refer to the groups with $b/a < 0.55$, the left panel of Figure \ref{fig:sample_cut}) and ``face-on'' (referring to $b/a \geq 0.55$, the right panel of Figure \ref{fig:sample_cut}). The edge-on groups have median $b/a \approx 0.4$, while the face-on groups have median $b/a \approx 0.7$.

In summary, we have eight groups, where each of the 4 bins in the SFR-M$_*$ plane are divided into edge-on and face-on axis ratio groups. We note that our results are unaffected by the choice of SFRs for this division. For example, we reach similar conclusions when using SED SFRs rather than \halpha SFR. This result is not surprising, as several works have shown that the SFRs agree well \citep[e.g.,][]{steidel_strong_2014, shivaei_mosdef_2016, theios_dust_2019, price_mosdef_2020}.

\begin{figure*}[tp]
\vglue -3pt
\centering
\includegraphics[width=0.95\textwidth]{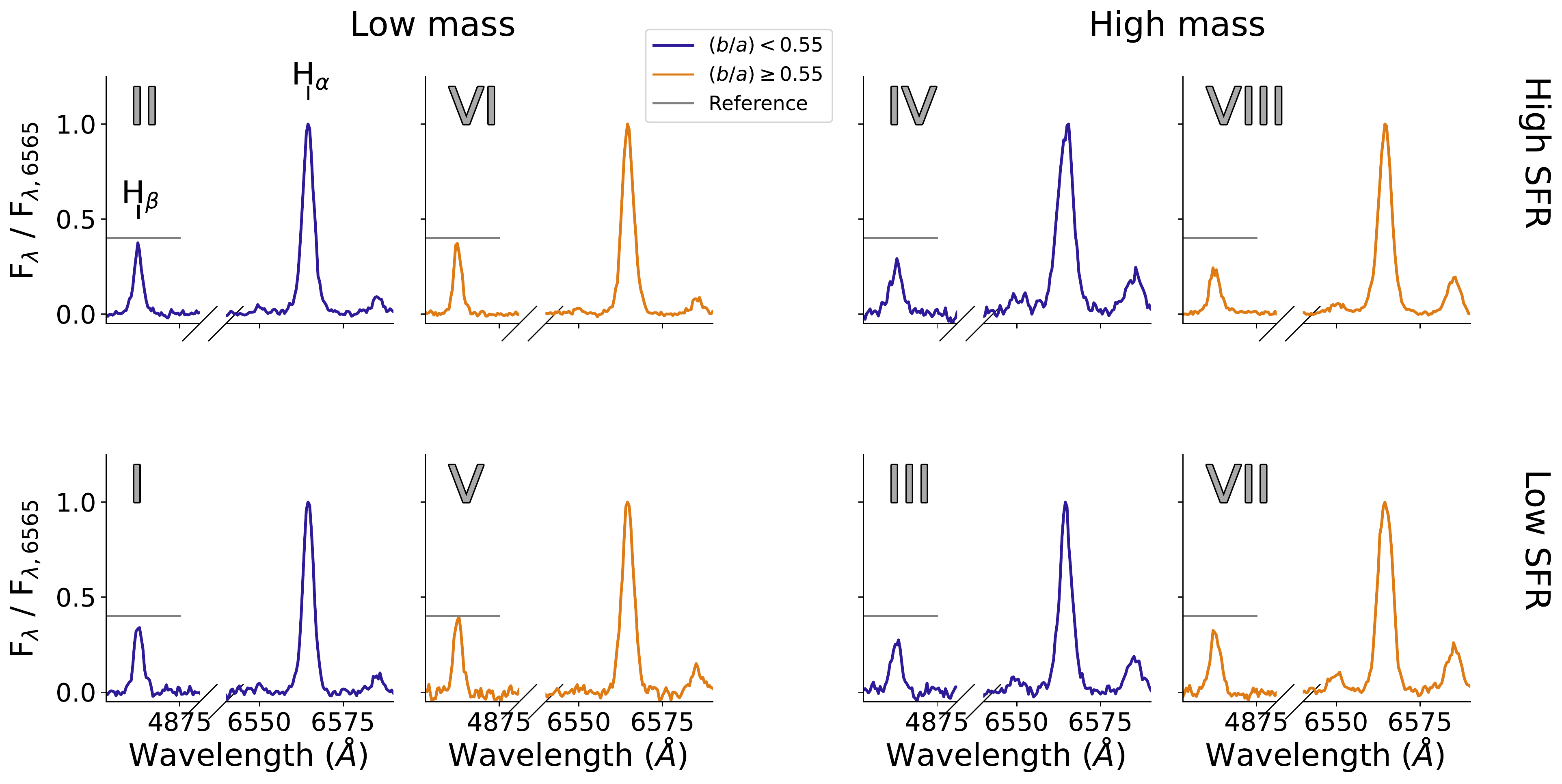}
\vglue -3pt
\caption{Stacked spectra for the eight groups illustrated in Figure \ref{fig:sample_cut}. We organize the spectra by stellar mass (left: low-mass, right: high-mass), SFR (bottom: low SFR, top: high SFR), and axis ratio (blue: edge-on, orange: face-on). The continuum has been subtracted for all stacked spectra. The spectra are normalized so they have the same \halpha peak flux for visual comparison. It is visible that the lower mass galaxies (left) have higher normalized \hbeta fluxes than the high-mass galaxies (right), which results in a smaller Balmer decrement for the lower mass galaxies. We do not observe a clear change in Balmer decrement with changing axis ratio for any of the groups.  
}
\label{fig:overlaid_spectra}
\end{figure*}

\begin{figure*}[!tp]
\vglue -5pt
\centering
\includegraphics[width=0.75\textwidth]{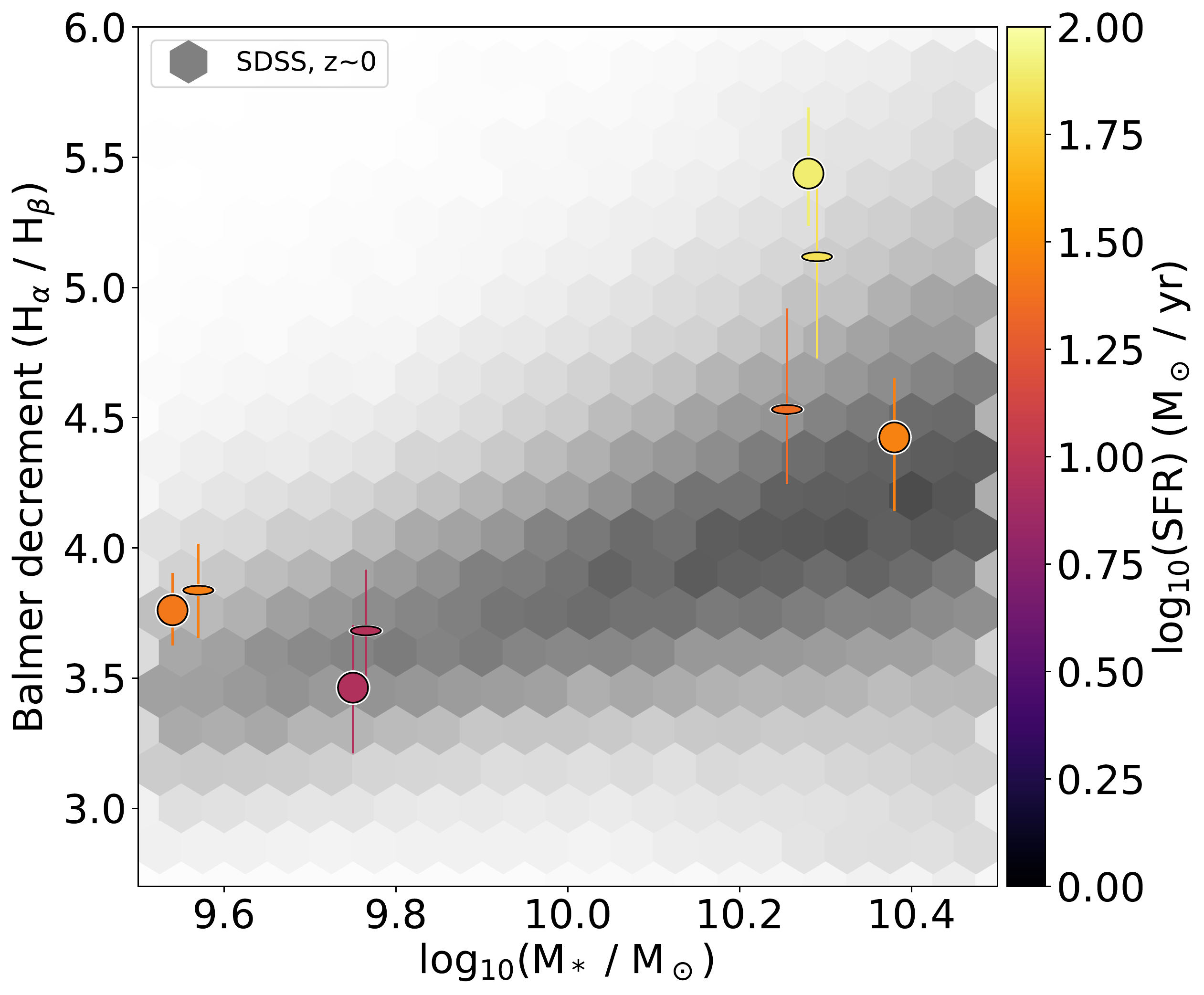}
\caption{
Balmer decrement vs. the median stellar mass in each of the eight groups color-coded by SFR, with the Balmer decrement of low-redshift SDSS galaxies shown in the grey background hexagons. Flat colored symbols show edge-on galaxies ($b/a<0.55$), while round symbols show face-on galaxies ($b/a\geq 0.55$). Edge-on galaxies have the same Balmer decrement as face-on galaxies 
at fixed mass and SFR in this sample. Consistent with previous findings, a clear trend with mass is visible, where higher mass galaxies have higher Balmer decrements \citep[e.g.,][]{garn_predicting_2010, shapley_mosfire_2022}. There does not appear to be strong residual trends with axis ratio or SFR for low-mass galaxies. At high mass, higher SFR galaxies have slightly higher Balmer decrements.}
\label{fig:balmer_mass_solo}
\end{figure*}

\subsection{Spectral Stacking} \label{subsec:spectral_stacking}

In order to substantially increase the signal-to-noise ratio of the spectra to measure an accurate Balmer decrement for a complete sample of galaxies, we employ spectral stacking within each of the eight groups.

First, we move the spectra of all galaxies to rest-frame and correct for the luminosity distance. As all MOSDEF spectra were flux calibrated, the resulting spectra present the luminosity density. Therefore, stacking in this manner effectively weighs the spectra by SFR, as a galaxy with higher SFR will have more flux in its Balmer lines and thus will contribute more to the Balmer decrement of the stack. In essence, this weighs the spectra such that each star-forming region within these galaxies are counted roughly equally. Therefore, studying these stacks gives insight into how dust affects star-forming regions at high redshift.

To compute a stacked spectrum, we first take the rest-frame spectrum of each galaxy in a group and create a mask to identify sky lines and regions with poor atmospheric transmission. We mask pixels where the measured error is at least 10 times larger than the median measured error of the spectrum. Next, all galaxies must be on the same wavelength axis to perform the stacking, so we linearly interpolate the spectra to 0.5\AA\ per pixel which is roughly the dispersion of the spectra in the rest-frame. Finally, before stacking, we extend the sky line mask to encompass one additional pixel to either side of the sky lines, as these pixels are far more uncertain due to the interpolation.

% Then, we normalize the galaxies such that the flux in their \halpha lines, as measured in \citet{reddy_mosdef_2015}, is the same value for all galaxies. This scaling ensures that each galaxy is contributing equally to the stack, so galaxies with strong \halpha emission or lower redshift do not wash out the galaxies in the group with lower \halpha fluxes. 

With all galaxies on the same wavelength axis and bad pixels masked out, we perform the stacking: at each pixel, we set the value of the stack to the \textit{median} of all of the non-masked points contributing to it. Due to the different redshifts of the galaxies, this procedure results in all of the galaxies contributing to the stacks in the regions of $4800-5100$\AA\ and $6400-6700$\AA\ rest-frame, but as low as $10\%$ contributing to the regions between these bands, such as $5500$\AA. For the purposes of this work, we are interested the emission line regions, so the larger uncertainties in the regions far from the lines are not relevant. Uncertainties on the spectrum are computed by adding the weighted errors of each contributing point in quadrature.

One concern of this technique is how well the emission lines from galaxies with different velocity dispersions will combine. To assess this potential issue, we generated synthetic \halpha and \hbeta lines with a realistic range of Balmer decrements ($3-6$), velocity dispersions ($70-150\, {\rm km\ s^{-1}}$), and redshifts ($1.6\leq z\leq 2.6$). Using the same stacking technique that we apply to the data, we stacked these synthetic lines and attempted to recover their mean Balmer decrements, changing several parameters to assess the recovery accuracy. First, increasing the resolution of the spectral interpolation from 0.5\AA\ to 0.001\AA\ had no effect on recovered Balmer decrements, so the effects of Nyquist bias seem negligible. Then, we tested smoothing the synthetic emission lines to the same velocity dispersion and observed no effect on the recovered Balmer decrements. In all cases, stacking using the median was better at recovering the sample median Balmer decrement, whereas stacking using the mean typically underestimated the median Balmer decrement by roughly $2\%$. As a result of these tests, we proceeded with the spectral stacking by taking the median of all pixels that contribute to a given point. However, we also re-ran the analysis in this paper with \textit{mean} stacking, and found similar results that are consistent within the uncertainties.

We also stack the FAST stellar continuum models using the same method as the spectra. We interpolate the models to the same wavelength axis as the spectra. Then, at each point, we take the median of all of the scaled FAST models that contribute to it. These continuum models include Balmer absorption, and are therefore necessary for deriving accurate measurements of the Balmer line fluxes in the stacks.

Since the FAST continuum models were generated from photometry, they need to be scaled to match the spectra. Assuming that the continuum has the correct shape, we compute two scale factors: one for the emission lines near \halpha (using $6300\mathrm{\AA}<\lambda<6900\mathrm{\AA}$) and one for the lines near \hbeta (using $4600\mathrm{\AA}<\lambda<5200\mathrm{\AA}$). When scaling the continuum, we mask out a broad region over the emission lines. Then, in each group, we subtract the scaled continuum model from the spectrum, leaving just the emission lines. We present a subset of the continuum-subtracted \halpha and \hbeta regions for the stacked spectra in Figure \ref{fig:overlaid_spectra}.

\subsection{Emission Fitting}

For each of the stacks, we fit \hbeta $4863$\AA, \OIII $4960$\AA\ and $5008$\AA, \halpha $6565$\AA, and \NII $6550$\AA\ and $6585$\AA. To first order, the lines are Gaussian, so all six emission lines are fit simultaneously with a Gaussian located at the expected center of each line at a fixed galaxy rotation velocity. We fit for redshift (which should be nearly zero, since these are rest-frame), velocity dispersion, and the amplitude of each line. With these fluxes, we compute Balmer decrements and metallicities for each group.

To obtain uncertainties on all of the emission line measurements and resulting properties (i.e., Balmer decrement and metallicity), we use bootstrapping. Within each group, we randomly draw $n$ galaxies with replacement, where $n$ is the size of the group. Then we stack their spectra as described in section \ref{subsec:spectral_stacking}, and repeat this process 100 times. Each of the 100 bootstrapped spectra in each group are fit and integrated with the same method as above, then a lower and upper $1\sigma$ uncertainty is computed as the $16$th and $84$th percentile of the resulting distribution, respectively.

% To obtain uncertainties on all of the emission line measurements and resulting properties (i.e., Balmer decrement and metallicity), we use Monte Carlo simulations, randomly perturbing each of the points on the spectrum 100 times then re-fitting the emission lines for each one. Taking the 16th and 84th percentiles of the resulting distribution provides lower and upper uncertainties on emission properties.

\subsection{Metallicity Measurement} \label{subsec:metallicity_measure}

Since gas-phase metallicity is an important parameter for understanding dust in galaxies, we also measure metallicities for the eight groups. Not all galaxies in the stacks cover the \OII 3726-3729\AA\ doublet, so we are limited in options for our metallicity calibration. Fortunately, we have all the required lines for O3N2, which is one of the more robust calibrators since it is computed from multiple line ratios and independent of dust effects \citep[e.g.,][]{liu_metallicities_2008, steidel_strong_2014, sanders_mosdef_2015}. Metallicities for the stacks are computed from this O3N2 line ratio, defined as

\begin{equation}
    \mathrm{O3N2} = \log_{10}\left(\frac{\mathrm{\OIII 5008\AA / \hbeta}}{\mathrm{\NII 6585\AA / \halpha}}\right).
\end{equation}

 We convert the measured O3N2 values to metallicity from the relation described in \citet{bian_direct_2018}, derived from observations of galaxies at low-redshift that are analogs of high-redshift galaxies. This relation is:

\begin{equation}
    12+\log_{10}\left(\mathrm{O/H}\right) = 8.97-0.39\times \mathrm{O3N2}.
\end{equation}

We also measure metallicities of the 100 bootstrapped spectra in each group, fitting their emission lines and using the same calculation. Uncertainties on the metallicities are then derived from the 16th and 84th percentile of the resulting measured metallicity distribution.

% We also measure metallicities of the 100 perturbed spectra in each group, fitting their emission lines and using the same calculation. Uncertainties on the metallicities are then derived from the 16th and 84th percentile of the resulting measured metallicity distribution.

\section{Results}
\label{sec:results}

With measurements of the eight stacked spectra, each constructed from galaxies that share similar mass, SFR, and axis ratio, we can now assess the effects of inclination on dust properties. 

First, we examine the relationship between dust, as measured by the Balmer decrement, and the mass, SFR, and axis ratios of the groups. In Figure \ref{fig:balmer_mass_solo}, we plot the measured Balmer decrements against median mass, color-coded by median SFR and shaped by median axis ratio of the galaxies in the group. Interestingly, edge-on galaxies have the same Balmer decrement as face-on galaxies at fixed mass and SFR. We recover the previously found trend that Balmer decrement increases with increasing stellar mass \citep[e.g.,][]{garn_predicting_2010, price_direct_2014, shapley_mosfire_2022}. For the low-mass galaxies, there does not appear to be any secondary trends between Balmer decrement and SFR. However, at high mass, higher SFR leads to a slightly higher Balmer decrement. 

\begin{figure*}[tp]
%\vglue -9pt
\centering
\includegraphics[width=0.95\textwidth]{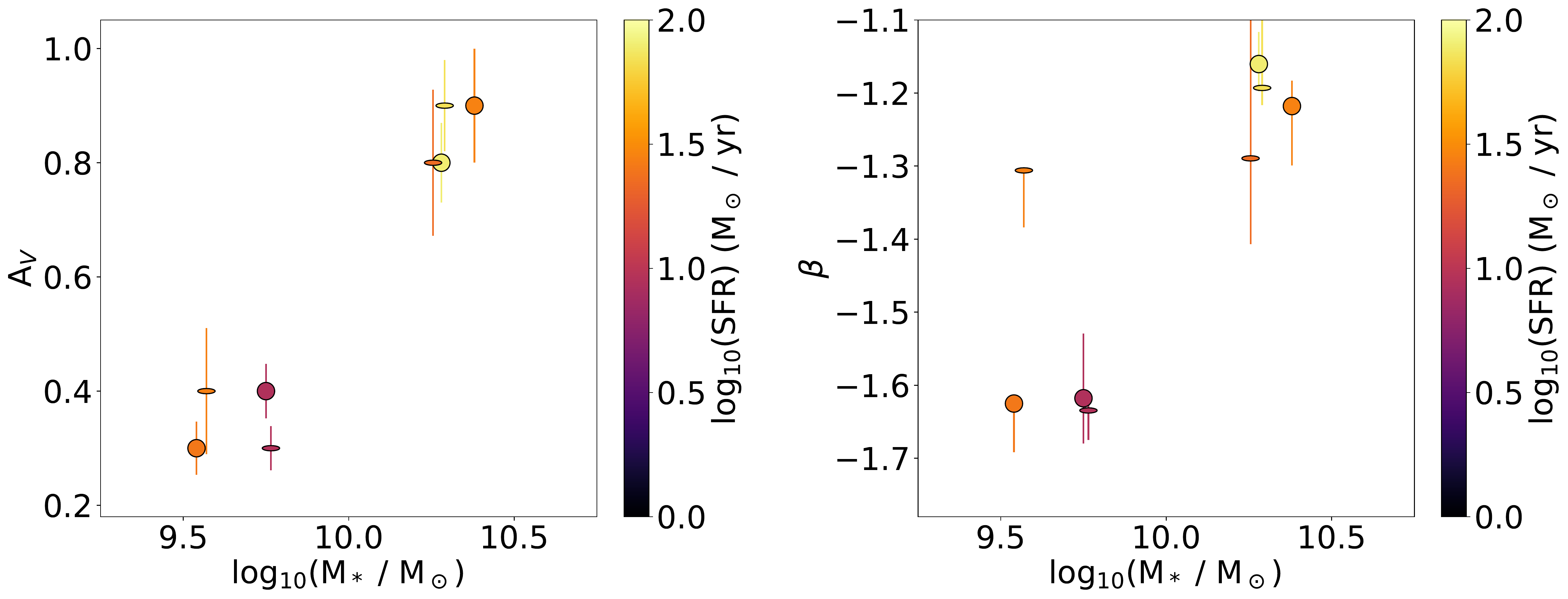}
%\vglue -8pt
\caption{
Attenuation in the V-band vs.\ stellar mass (left) and UV slope ($\beta$) vs.\ stellar mass (right). Both \AV and $\beta$ are measured as the median of fits to the SEDs of the individual galaxies in each group. \AV and $\beta$ appear to be similar for edge-on and face-on galaxies at fixed mass and SFR. We see a strong correlation in both \AV and $\beta$ with mass, but no residual trends with axis ratio or SFR. 
}
\label{fig:av_beta_combined}
\end{figure*}

Each galaxy also has a measured attenuation in the V-band (\AV) and UV slope ($\beta$) from the best-fit stellar population model to their SED (see Section \ref{subsec:mosdef_survey}). For each of the groups, we plot their median \AV and $\beta$ values vs. stellar mass, colored by SFR and shaped by axis ratio in Figure \ref{fig:av_beta_combined}. Again, the dust properties of edge-on and face-on galaxies in our sample appear to be similar at fixed mass and SFR. We observe a strong trend of increasing dust content with increasing mass. There is no apparent secondary dependence on SFR or axis ratio.

\begin{figure}[t]
\centering
\includegraphics[width=0.49\textwidth]{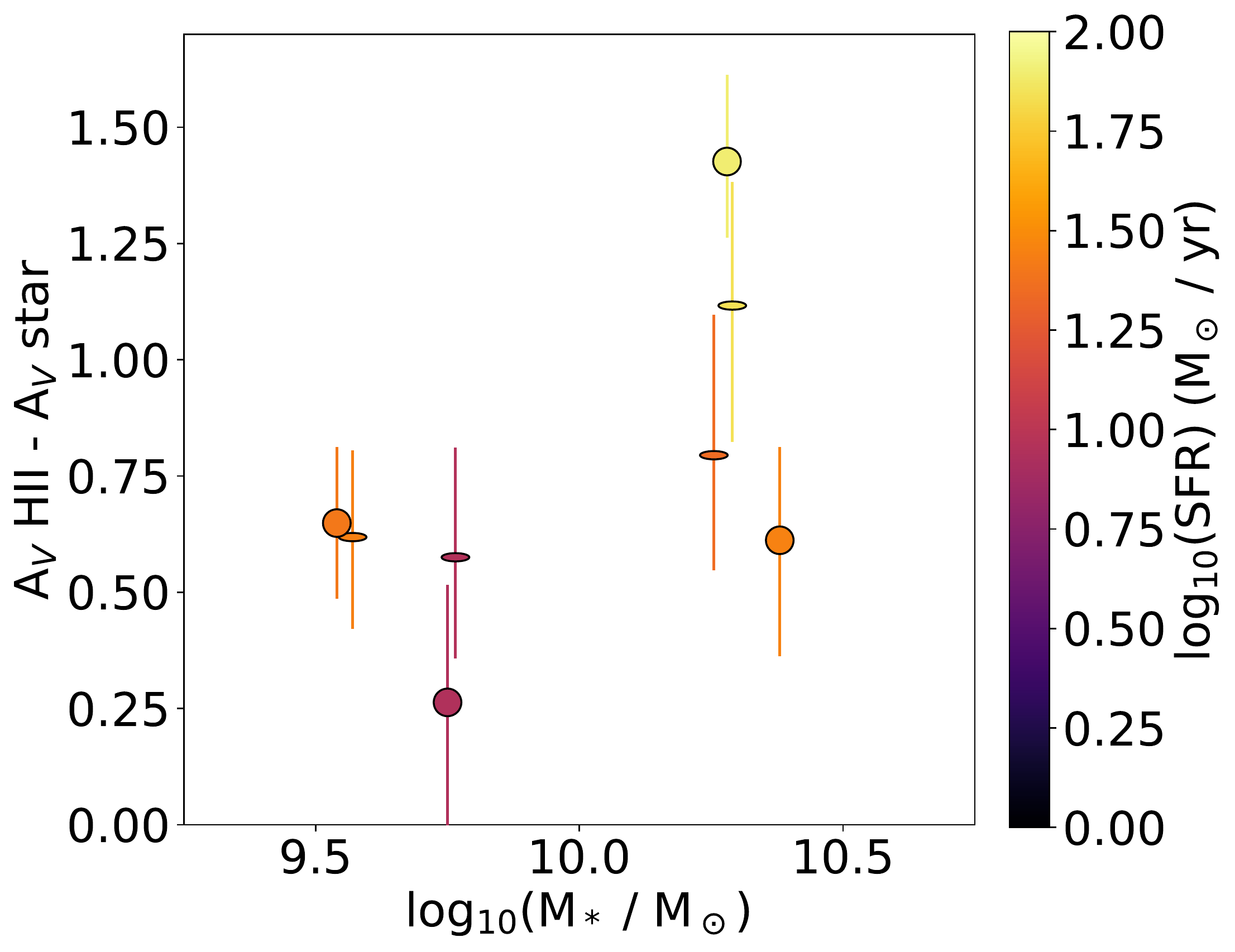}
\caption{The difference in attenuation between the Balmer lines and stellar continuum, measured in the V-band, plotted against stellar mass. Again, we see no effects with inclination. We also observe that galaxies with higher mass and SFR tend to have higher excess nebular attenuation than galaxies at lower masses.}
\label{fig:av_extra_mass}
\end{figure}

We also investigate the relationship between attenuation of the Balmer lines and attenuation of the stellar continuum. The difference between these values, calculated in the V-band, is displayed in Figure \ref{fig:av_extra_mass} as a function of stellar mass. Consistent with our previous results, there is no correlation with inclination. However, we do observe that galaxies with higher mass and SFR have increasingly more attenuation in the Balmer lines than the stellar continuum, which places an additional constraint on the dust geometry (see Section \ref{subsec:implications_dust}). Similar results of increasing Balmer attenuation relative to continuum with increasing mass are observed in for distant star-forming galaxies by \citet{price_direct_2014, reddy_mosdef_2020}.

\begin{figure*}[t]
\centering
\includegraphics[width=0.66\textwidth]{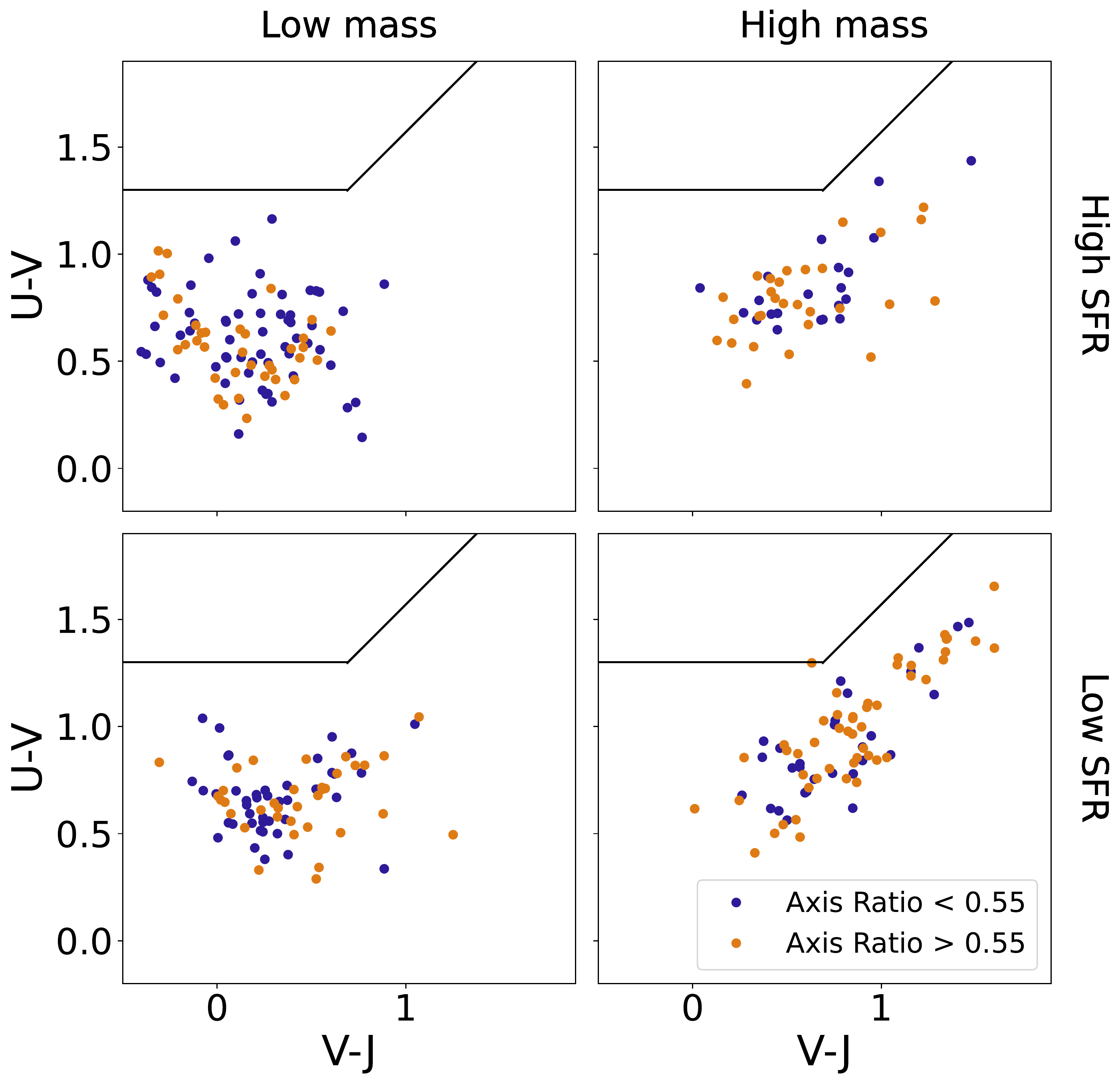}
\caption{
The UVJ diagram for each of the eight groups, organized by stellar mass (left: low-mass, right: high-mass), SFR (bottom: low SFR, top: high SFR), axis ratio (blue: edge-on, orange: face-on). All points fall in the region of star-forming galaxies on the diagram. Galaxies with more dust tend to be located towards the upper-right of the diagram \citep[e.g.,][]{patel_uvj_2012}. Within each group, the distribution of points is similar, indicating that there is no correlation between dust and axis ratio. Between groups, the high-mass galaxies appear to have higher U-V and V-J which is consistent with high-mass galaxies being more dusty. 
}
\label{fig:uvj_ar_groups}
\end{figure*}

To further search for any trends with axis ratio, we plot all galaxies in each group on a UVJ diagram (Figure \ref{fig:uvj_ar_groups}), separated by mass and SFR. The UVJ diagram is a useful diagnostic, as more dusty galaxies tend to be located towards the upper-right of the diagram \citep[e.g.,][]{patel_uvj_2012}. Within each mass and SFR bin, the distributions for low axis ratio and high axis ratio galaxies are similar. The lack of visible trends with axis ratio in the UVJ diagram further confirms our results of no trend between dust properties and galaxy inclination. Again, we recover a strong trend with mass, where the higher mass galaxies tend to be at higher values of both U-V and V-J, indicating higher dust attenuation \citep[e.g.,][]{labbe_irac_2005,williams_detection_2009, brammer_dead_2009}.

In summary, we observe no trends between Balmer decrement, \AV, $\beta$, and UVJ diagram position with axis ratio. We conclude that dust properties appear to be independent of axis ratio at the regime of redshift, mass, and SFR probed by the MOSDEF survey. We also find that Balmer decrement, \AV, $\beta$ increase with increasing galaxy mass. Additionally, there is more nebular attenuation than stellar attenuation, and this difference grows larger with increasing galaxy mass and SFR. A table of our measured values for mass, SFR, axis ratio, and dust properties is displayed as Table \ref{tab:axis_ratio_data}.

\begin{table*}[t]
  \centering
  \vspace{0.2cm}  
  \begin{tabular}{ccccccccc}\hline\hline
    $\rm{Group}$ & $\overline{\log_{10}(\rm{M}^*)}\ (M_\odot)$ & $\overline{\log_{10}(\rm{SFR})}\ (M_\odot \rm{yr}^{-1})$ & $\overline{\rm{Axis\ Ratio}}$ & $\overline{\rm{z}}$ & $12 + \log_{10}{\left(\rm{O}/\rm{H}\right)}$  & $\overline{\rm{\AV}}$ &  $\overline{\rm{\beta}}$ & $\left(\halpha/\hbeta\right)$ \\
    \hline
    I & $9.77$ & $0.96$ & $0.38$ & $2.2$ & $8.36\pm_{0.03}^{0.03}$ & $0.3\pm{0.04}$ & $-1.63\pm_{0.04}^{0.05}$ & $3.68\pm_{0.2}^{0.23}$  \\ 
II & $9.57$ & $1.44$ & $0.41$ & $2.23$ & $8.32\pm_{0.0}^{0.05}$ & $0.4\pm{0.12}$ & $-1.31\pm_{0.08}^{0.0}$ & $3.84\pm_{0.18}^{0.18}$  \\ 
III & $10.26$ & $1.35$ & $0.42$ & $1.92$ & $8.61\pm_{0.03}^{0.0}$ & $0.8\pm{0.12}$ & $-1.29\pm_{0.12}^{0.24}$ & $4.53\pm_{0.29}^{0.39}$  \\ 
IV & $10.29$ & $1.84$ & $0.42$ & $2.23$ & $8.55\pm_{0.01}^{0.02}$ & $0.9\pm{0.08}$ & $-1.19\pm_{0.02}^{0.11}$ & $5.12\pm_{0.39}^{0.38}$  \\ 
V & $9.75$ & $0.94$ & $0.69$ & $2.11$ & $8.44\pm_{0.03}^{0.03}$ & $0.4\pm{0.05}$ & $-1.62\pm_{0.05}^{0.06}$ & $3.46\pm_{0.25}^{0.24}$  \\ 
VI & $9.54$ & $1.41$ & $0.72$ & $2.2$ & $8.31\pm_{0.01}^{0.02}$ & $0.3\pm{0.06}$ & $-1.63\pm_{0.05}^{0.03}$ & $3.76\pm_{0.14}^{0.14}$  \\ 
VII & $10.38$ & $1.45$ & $0.7$ & $2.18$ & $8.68\pm_{0.04}^{0.0}$ & $0.9\pm{0.13}$ & $-1.22\pm_{0.11}^{0.03}$ & $4.42\pm_{0.28}^{0.23}$  \\ 
VIII & $10.28$ & $1.9$ & $0.71$ & $2.27$ & $8.58\pm_{0.01}^{0.01}$ & $0.8\pm{0.07}$ & $-1.16\pm_{0.07}^{0.04}$ & $5.44\pm_{0.2}^{0.25}$  \\ 
\hline
\end{tabular}
\caption{\label{tab:axis_ratio_data} For each of the eight groups, we display our measured values of: stellar mass, SFR, axis ratio, redshift, metallicity, \AV, $\beta$, and Balmer decrement. A bar over the symbol denotes that the property is a median of the constituent galaxies in a group.}
\end{table*}

\section{Discussion}
\label{sec:discussion}

In this section, we propose explanations for why the Balmer decrement, \AV, and $\beta$ do not depend on galaxy axis ratio and what this implies for the dust geometry (Section \ref{subsec:implications_dust}). We then compare this work to past studies (Section \ref{subsec: compare_studies}). Finally, we demonstrate why at fixed mass the Balmer decrement might exhibit no secondary dependence with SFR or metallicity, as seen in the low-mass galaxies in our sample (Section \ref{subsec:dust_model}).

\subsection{Implications for Dust-to-Star Geometry}
\label{subsec:implications_dust}

The optical depth of dust depends on the opacity and density of the obscuring material, as well as the path length of the light through the galaxy. Thus, we may expect that dust attenuation would correlate with axis ratio, as the light from more edge-on galaxies may encounter more obscuring material. However, we find that the Balmer decrement does not depend on the inclination, implying that light from the youngest stars in edge-on and face-on galaxies encounter the same amount of obscuration. 

Interestingly, similar to the Balmer decrement, \AV and $\beta$ also do not change with inclination. \AV and $\beta$ measure the attenuation towards all stars, not just the young stellar population. Typically, the older stellar population is only attenuated by the diffuse ISM, as older stars are no longer in star-forming regions. Thus, we would expect to see a clear dependence of \AV and $\beta$ with axis ratio. Our finding that there is no trend between Balmer decrement, \AV, or $\beta$ with axis ratio seems to contradict any model where ISM dust plays a large role in attenuation. For example, the two-component model mentioned in the introduction does not appear to hold, as it predicts a substantial ISM dust component which would lead to inclination-dependent attenuation. However, the galaxies in our sample have relatively high sSFRs, even below the star-forming main sequence ($-9.0 < \log_{10}\left(\mathrm{sSFR}\ yr^{-1}\right) < -8.1$). Therefore, a substantial fraction of the stellar continuum light from these galaxies may be emitted from young stars in star-forming regions rather than the older stellar populations, and ISM dust might not significantly attenuate the stellar light.

For a dust model to be consistent with our results, the model must be able to explain the following findings. \begin{enumerate}
    \item Dust attenuation for both the Balmer lines and stellar continuum do not depend strongly on inclination (Figures \ref{fig:balmer_mass_solo}, \ref{fig:av_beta_combined}).
    \item The total amount of attenuation in the Balmer lines and continuum increases with galaxy mass (Figures \ref{fig:balmer_mass_solo}, \ref{fig:av_beta_combined}).
    \item The Balmer lines are increasingly more attenuated than the continuum as galaxy mass and SFR increase (Figure \ref{fig:av_extra_mass}). 
\end{enumerate} 

Based on these constraints, we propose a dust model that has attenuation occurring in three main components: large star-forming clumps, typical star-forming regions, and a very small amount from the ISM \citep[see also][]{reddy_mosdef_2020}. This model is similar to the two-component model found to be valid at low redshift, with the addition of large, dusty, star-forming clumps. 

Large star-forming clumps are indeed observed in galaxies at $z\sim 2$ in \halpha and rest-frame UV maps. These large, 1kpc-scale  structures can contribute up to 20\% of the star formation in the galaxy \citep{wuyts_smoother_2012, schreiber_constraints_2011}. Also, as galaxy mass increases, clumps tend to be larger \citep{swinbank_properties_2012} and more common \citep{tadaki_nature_2014}. As they are detected through \halpha, these clumps are a strong source of Balmer emission. 

\begin{figure*}[th]
\centering
\includegraphics[width=0.66\textwidth]{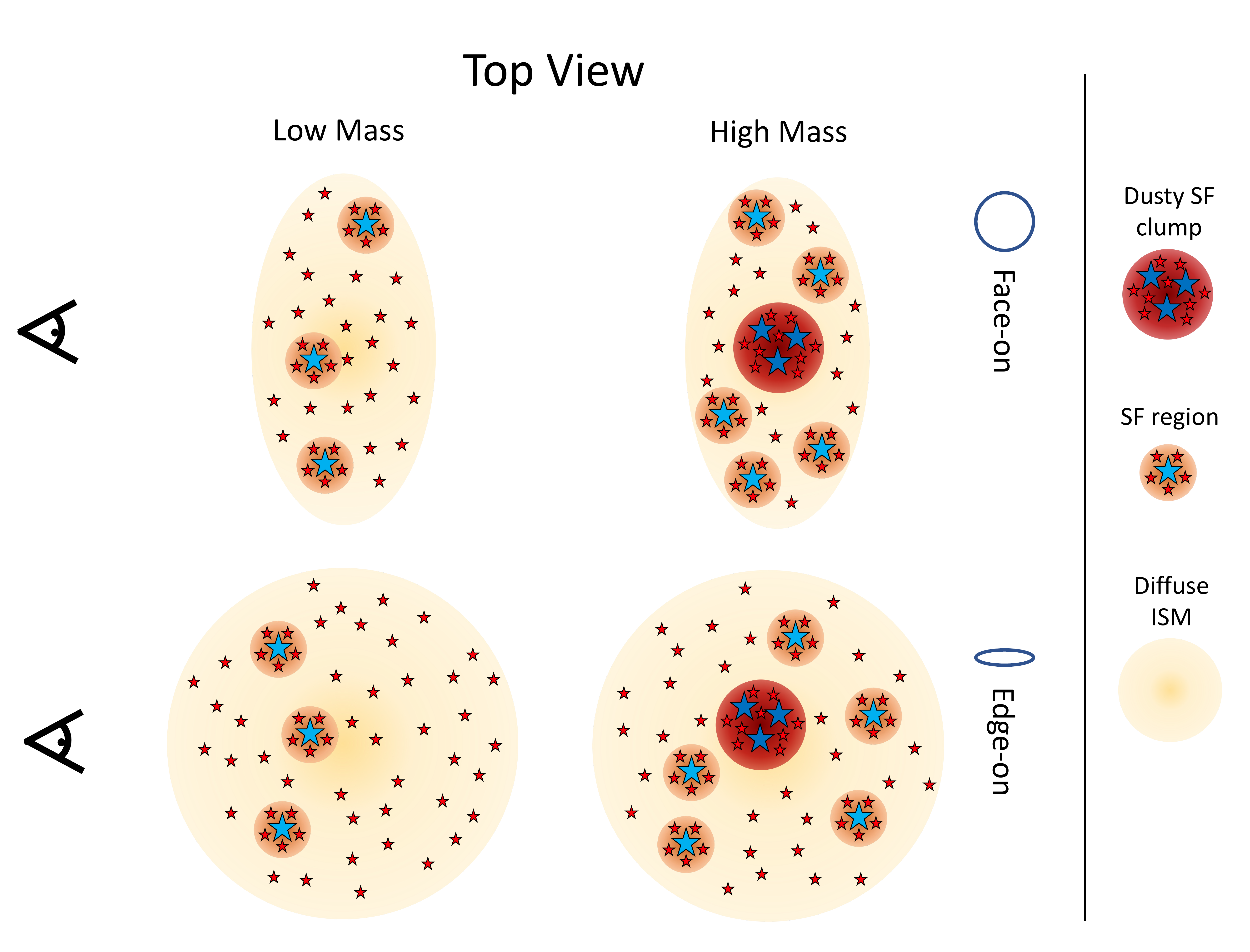}
\vglue -10pt
\caption{
One low-mass galaxy (left) and one high-mass galaxy (right) viewed by an observer as a face-on (top) and edge-on (bottom) system. The observer is represented by an eye, and the reader is viewing the picture from above. The yellow background represents dust in the diffuse ISM, the orange represents dust in typical-size star-forming regions, and the red represents dust in the large 1kpc-scale star-forming clumps. Both of the star-forming structures are populated with young stars, but there is a higher fraction of young stars in the clumps. In all cases, since the ISM is not contributing much to attenuation, the observer sees similar attenuation in both face-on and edge-on orientations. Also, at higher masses (right), larger attenuation is observed due to the longer average path-length of light through the larger clumps. Finally, the excess attenuation (A$_\mathrm{Balmer}-$A$_\mathrm{V}$) is larger for the high mass (high SFR) galaxies, due to the extra Balmer light originating in the massive clumps. 
}
\label{fig:dust_diagram_axisratio}
\end{figure*}

Our dust model meets all of the constraints imposed by the data. First, to meet the constraint of independence from inclination effects, the ISM is not a major contributor to attenuation in our model. The observed galaxies have high sSFRs, which causes a significant fraction of the stellar light to be coming form star-forming regions rather than the older population. Then, since most attenuation occurs in (roughly spherically symmetric) star-forming regions and clumps, changes in attenuation with inclination would not be observed (constraint 1). Second, clumps are larger and more common as galaxy mass increases. Since these clumps emit a fraction of the blue stellar light, as well as a large amount of Balmer emission, the average path length of the light increases with mass for both the continuum and Balmer lines (constraint 2). Third, Balmer attenuation is dominated by clumps while stellar attenuation primarily probes the less-obscured, typical, star-forming regions. Thus, the model predicts that the Balmer attenuation will be larger than continuum attenuation since clumps are more dusty. Then, as galaxy mass and SFR increase, the clump size and prevalence increases, and so the model predicts an increasing difference between Balmer and stellar attenuation (constraint 3). Our model meets all constraints of the data, and is presented pictorially in Figure \ref{fig:dust_diagram_axisratio}.

A different explanation for our results might be that dusty star formation in these galaxies is concentrated toward the center, perhaps in a bulge-like structure. There are some indications for this in high-redshift galaxies, where ALMA observations detected high central IR star formation rates in high mass galaxies at $z=2.5$ \citep[e.g.,][]{barro_sub-kiloparsec_2016}. If the dominant source of Balmer emission and attenuation is centrally located, and it occurs in a structure that is roughy spherical, then the encountered dust column density might not change substantially with inclination (constraint 1). In this central star-formation model, larger galaxies have larger regions of central star formation, and thus a longer path length through attenuating material to escape (constraint 2). Finally, to explain the larger differential attenuation between Balmer lines and continuum emission, the youngest stars must be formed closer to the center, while other star-forming regions might be located further out. As the galaxy mass increases, the fraction of star formation from the dusty central starburst increases, and thus the nebular attenuation increases faster than the stellar attenuation (constraint 3). 

Finally, it is expected that scattering of thermal emission in the disk of edge-on galaxies would lead to radiation loss in directions perpendicular to the disk \citep{misiriotis_modeling_2001}. We note that, for our sample, we also do not observe any difference in MIPS 24 micron emission between edge-on and face-on galaxies with similar properties. Though the MIPS 24 micron band does not probe thermal emission well since it is in the PAH regime, this result points to the fact that energy balance is similar for both orientations. The reason that we do not see any difference may again be due to the high sSFR of our sample, which might cause most dust obscuration to occur in spherically symmetric regions. If indeed dust spread throughout the whole disk doesn't play a large role in scattering thermal emission, there would not be an expected trend with axis ratio.  

In the future, our proposed geometry can be tested with dust measurements of individual galaxies across the full range of axis ratios and SFRs, as well as further investigation into the distribution of star-formation and clumps in high-redshift galaxies.

\subsection{Comparison to Other Studies}
\label{subsec: compare_studies}

Our data support a geometry where attenuating dust is mostly concentrated around star-forming regions and clumps, explaining the observations that Balmer decrement does not vary with axis ratio. The ISM does not seem to play a large role in attenuation, especially due to the relatively high sSFR of the galaxies. Our findings fit in the framework suggested by \citet{reddy_mosdef_2020}, in which young star-forming regions may be dustier and contribute strongly to nebular attenuation. Our observations are also partially in agreement with the patchy dust model proposed by \citet{reddy_mosdef_2015}, where the youngest stars become increasingly more obscured as SFR increases. However, this model is not fully consistent with our observations, since it also predicts that the stellar attenuation is dominated by the more diffuse ISM dust, and thus we would expect to see a trend between axis ratio and \AV or $\beta$. Another patchy dust model is presented in \cite{fetherolf_mosdef_2023}, where dust transitions from smooth to patchy as galaxy mass increases, and dusty clumps are predicted at high masses. This model is consistent with our results as long as the ISM dust is not responsible for much of the observed stellar attenuation.

% By a similar argument as for the \HII regions, it might be that any variation of \AV or $\beta$ with axis ratio is negligible unless the system is almost perfectly edge-on. Since very few galaxies in our sample would be at such low axis ratios, any effect on \AV or $\beta$ might be washed-out in the stacked spectra. 

The lack of any Balmer decrement correlation with axis ratio is consistent with \citet{yip_extinction_2010},  \citet{battisti_characterizing_2017}, and \citet{yuan_asymmetry_2021}, who all found analogous results for samples at similar stellar masses at low redshift, and consequently lower SFR. \citet{yip_extinction_2010} and \citet{yuan_asymmetry_2021} do observe that the stellar continuum attenuation is higher in low-redshift edge-on galaxies compared to face-on galaxies. Nonetheless, they conclude that dust attenuation primarily takes place in \HII regions, and attenuation in the ISM of the thick disk is insignificant, similar to our results. Our results also contrast the work by \citet{patel_uvj_2012} at lower redshift $\left(0.6<z<0.9\right)$ who do find that galaxies with higher $V-J$ values have lower axis ratios. Similar results were found by \citet{zuckerman_reproducing_2021} who show that the wide range of dust attenuation values measured for star-forming galaxies at a given redshift and stellar mass is almost entirely due to the effect of inclination. This result is mostly valid for $z<1.5$, but it is less apparent at higher redshifts. These inconsistencies may be partly explained by the lower sSFR at later times, as less of the stellar light would originate from  star-forming regions. Furthermore, there may be more ISM dust at lower redshifts. Indeed, at $z\approx 1.4$ and lower sSFR than this work, \citet{price_direct_2014} demonstrate a relationship between sSFR and the difference between nebular and stellar attenuation for star-forming galaxies in the 3D-HST survey. 

\subsection{Explaining why Balmer Decrement Depends Primarily on Mass} \label{subsec:dust_model}

In Section \ref{sec:results}, we found that Balmer decrement is independent of SFR at fixed stellar mass for the low-mass galaxies and shows only a minor dependence with SFR for the more massive galaxies. This result may seem puzzling since galaxies with higher SFRs at fixed mass have higher gas fractions \citep[e.g.,][]{kennicutt_global_1998}, and thus may be expected to have more dust. The amount of dust in a galaxy is connected to the amount of gas through gas-phase metallicity. Thus, in order to explain our results, the next natural step is to investigate the metallicities of our galaxies. 

\begin{figure}[tp]
\centering
\includegraphics[width=0.45\textwidth]{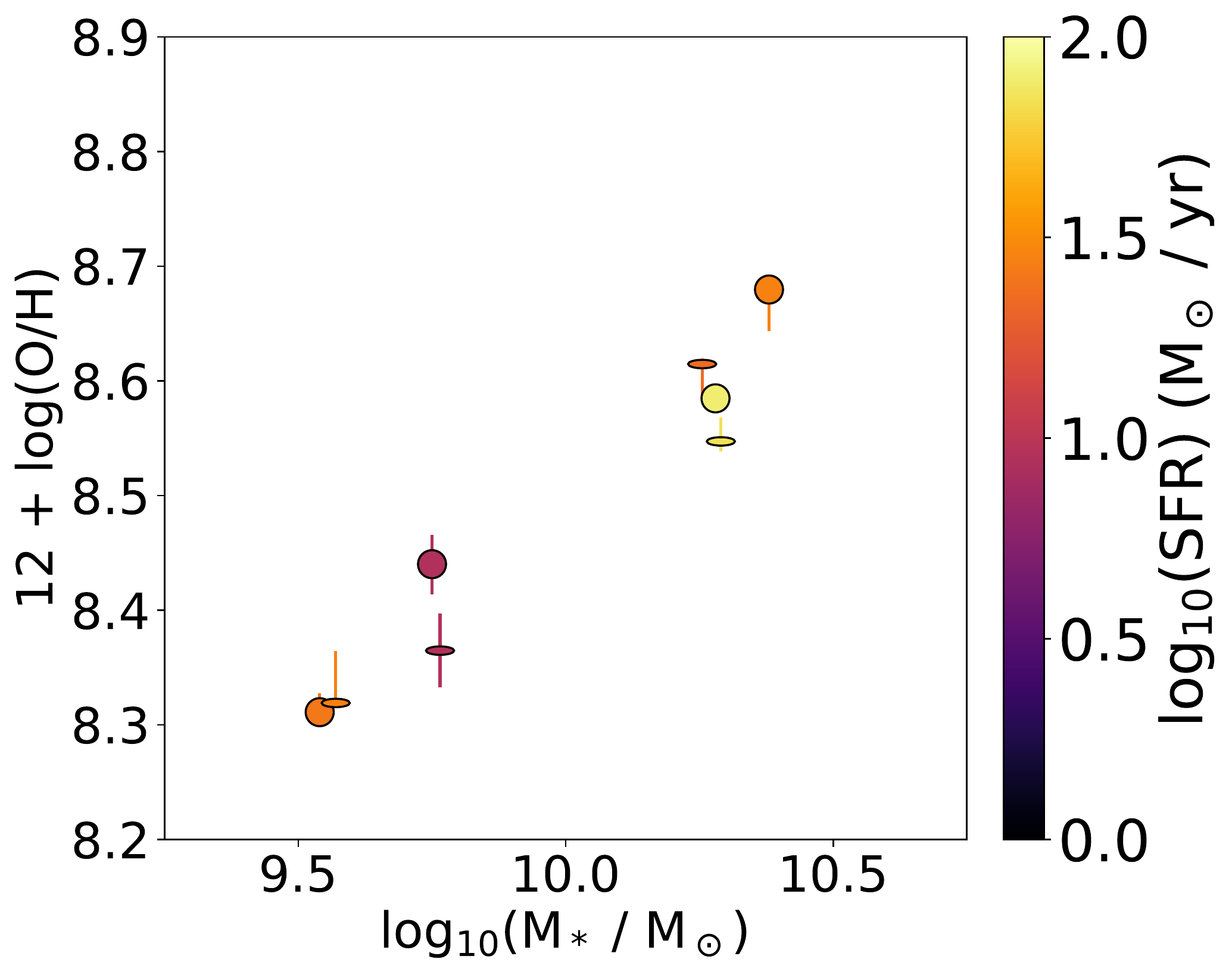}
\caption{
Gas-phase metallicity computed from O3N2 plotted against stellar mass for each of the eight groups, where points are colored by SFR. The measured metallicities show a clear mass-metallicity relation as expected. At similar stellar mass, higher SFR leads to lower metallicity, also as expected.
}
\label{fig:mass_metallicity}
\end{figure}

In Figure \ref{fig:mass_metallicity} we show that the median mass of each group and measured metallicity (from stacked spectra) recovers the well-known mass-metallicity relation \citep{tremonti_origin_2004}: higher mass galaxies have larger gravitational potentials, so they retain more metals. Figure \ref{fig:mass_metallicity} also shows that we recover the Fundamental Metallicity Relation (FMR) \citep{mannucci_fundamental_2010}: higher SFR leads to lower metallicity at fixed mass. This anti-correlation between SFR and metallicity is thought to arise because high SFR tends to be indicative of high inflow of metal-poor gas from the intergalactic medium, and vice versa. We refer to \citet{sanders_mosdef_2015, sanders_mosdef_2018, sanders_mosdef_2020, sanders_mosdef_2021} for detailed studies on the FMR in the MOSDEF survey.

\begin{figure*}[!tp]
\centering
\includegraphics[width=0.95\textwidth]{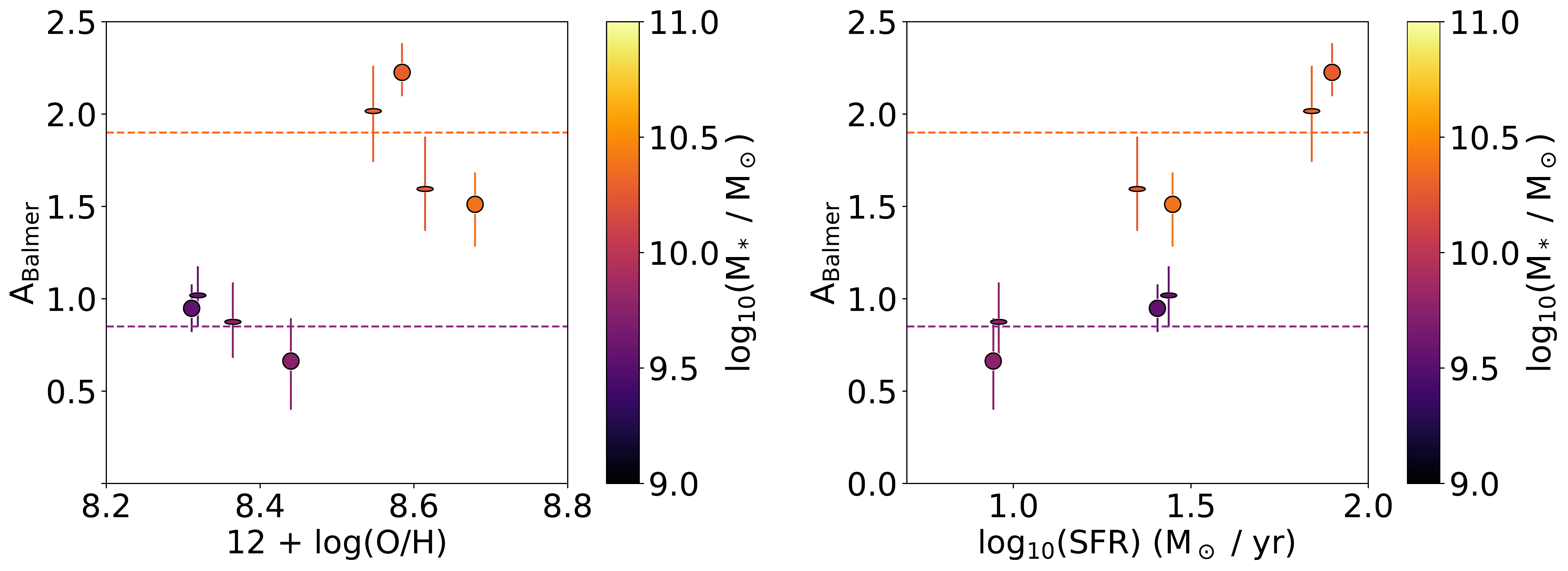}
\caption{
Balmer attenuation vs.\ metallicity (left) and vs.\ SFR (right), colored by median stellar mass. Dashed lines of the average Balmer attenuation for the four low-mass (purple) and four high-mass (orange) groups are shown. At similar Balmer attenuation and stellar mass, we observe a wide range of SFRs and metallicities for the low-mass galaxies. At high mass, there appears to be a small trend where Balmer attenuation decreases with increasing metallicity, and increases with increasing SFR. We conclude that Balmer attenuation primarily depends on stellar mass for the low-mass galaxies in our sample.}
\label{fig:balmer_sfr_metallicity}
\end{figure*}

\begin{figure}[h]
\centering
\includegraphics[width=0.5\textwidth]{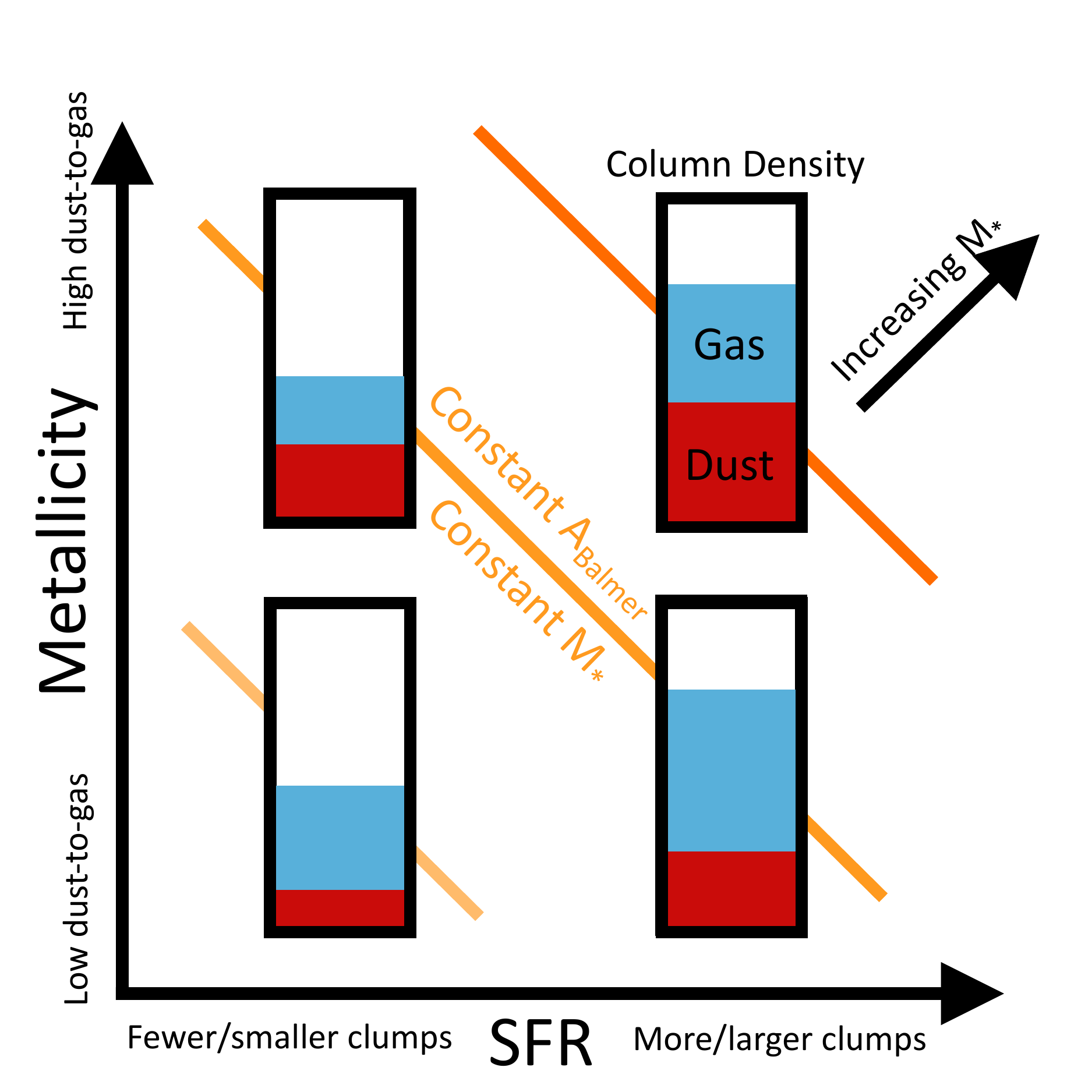}
\vglue -3pt
\caption{A diagram depicting how the relationship between SFR and metallicity affects Balmer attenuation. The bars represent the column density of gas (blue) and dust (red) encountered by escaping light for the galaxies that fall in the corresponding parts of the SFR-metallicity plane. Lines of roughly constant A$_\mathrm{Balmer}$ (from Equation \ref{eq:dust_model}) and constant mass (from the FMR) are shown in the background in orange. Galaxies at higher SFR (right) have more star-forming clumps, so the light sees a longer path-length of gas and dust on average. Galaxies with higher metallicities (top) have higher dust-to-gas ratios, so a larger fraction of their gas column density is filled with obscuring dust. This representation shows that galaxies with the same dust column density (i.e. top-left and bottom-right) and thus same Balmer attenuation can have very different metallicities and SFRs.
}
\label{fig:dust_diagram_metallicity}
\end{figure}

Since Balmer decrement increases with mass, we might expect from the mass-metallicity relation that Balmer decrement would also increase with metallicity. While this trend holds at a large-scale level, our data reveal that the Balmer attenuation does not strongly depend on metallicity at fixed mass. The left panel of Figure \ref{fig:balmer_sfr_metallicity} shows that one value of Balmer attenuation (calculated directly from the Balmer decrement as in \citet{price_direct_2014}) corresponds to a wide range of metallicities for low-mass galaxies, spanning 0.2 dex. At high mass, there seems to be a trend in the opposite direction than expected, where higher metallicities are exhibiting lower Balmer decrements.  Similarly, we show that there is no correlation between Balmer attenuation and SFR for the low-mass galaxies (Figure \ref{fig:balmer_sfr_metallicity}, right panel). 

Our results of roughly constant Balmer attenuation at constant mass imply that these galaxies, despite the range in SFRs and metallicities, must have approximately the same dust column density. Fundamentally, this arises from the following argument. Assuming a fixed metal yield, the amount of metals in a galaxy only depends on the total mass formed. Since dust is formed from metals, we expect that the dust column density (and thus the Balmer decrement) similarly only depends on the galaxy mass. On the other hand, since metallicity is defined with respect to the hydrogen content, metallicity depends on both the dust column density and the gas column density. For example, adding more low-metallicity gas (i.e. infalling from the intergalactic medium) will not change the dust column density, but it will lower the metallicity. Therefore, we find a range of metallicities at fixed Balmer decrement, which are due to changes in the gas fraction.

Another way to understand the result that Balmer decrement depends only on stellar mass is to use observed relations to examine the interplay between SFR and metallicity. To compare these properties, we consider both the Fundamental Metallicity Relation and the Kennicutt-Schmidt (K-S) Relation \citep{kennicutt_global_1998}. These relations together indeed provide a qualitative explanation for how such a range of SFRs and metallicities have the same Balmer attenuation. The K-S relation states that, at fixed mass, galaxies with higher SFR will have higher hydrogen gas content, fueled by infalling intergalactic medium (IGM) gas. This IGM gas is metal-poor, resulting in a decrease of metallicity as SFR increases --- this relationship is the FMR. By definition, a low metallicity implies a low dust-to-gas ratio. Therefore, galaxies with higher SFRs have more gas, but a lower dust-to-gas ratio, and thus can have the same dust column density as galaxies with low SFRs and consequently high dust-to-gas ratios. The relation between metallicity, SFR, and the Balmer decrement is shown in the Figure \ref{fig:dust_diagram_metallicity} diagram.

To assess whether these different relations indeed lead to a range of SFRs and metallicities at fixed Balmer decrement, we now explore this model quantitatively. As supported by our results, we assume a dust geometry that does not depend on inclination (e.g., Section \ref{subsec:implications_dust}). We assume that this attenuating dust has a dust mass absorption coefficient, $\kappa_\lambda$, which is spatially independent and that the dust is spread evenly throughout the attenuating region (either typical star-forming regions or dusty star-forming clumps). From these two assumptions, \citet{shapley_mosfire_2022} derived that dust attenuation is proportional to dust-to-gas ratio and gas surface density:
\begin{equation}\label{eq:alice_a_lambda}
    A_\lambda \approx 1.086\times \kappa_\lambda\times \left(\frac{M_{\mathrm{dust}}}{M_{\mathrm{gas}}}\right)\times \Sigma_{\mathrm{gas}}.
\end{equation}
We can replace $\left(M_{\mathrm{dust}} / M_{\mathrm{gas}}\right)$ using the power-law relation from \citet{de_vis_systematic_2019}, which is
\begin{equation}
    \log_{10}\left(\frac{M_{\mathrm{dust}}}{M_{\mathrm{gas}}}\right) = a\times \logoh + b.
\end{equation}
Empirically, they find $a\approx 2.15$ for the O3N2 metallicity calibration, and we can fold $b$ into our constants. We also note that there does not appear to be significant evolution in the dust-to-gas ratio from $z=0$ to $z\approx 2$, so this relation is applicable \citep{shapley_first_2020, popping_dust--gas_2022}. We can use the K-S Relation \citep{kennicutt_global_1998} to replace $\Sigma_{\mathrm{gas}}$ from Equation \ref{eq:alice_a_lambda}, given by
\begin{equation}
    \Sigma_{\mathrm{gas}} \propto \left(\Sigma_{\mathrm{SFR}}\right)^{\left(\frac{1}{n}\right)}.
\end{equation}
We now replace $\Sigma_{\mathrm{SFR}}$ with its definition, focusing specifically on the attenuating region, which is
\begin{equation}
    \Sigma_{\mathrm{SFR}} = \frac{\mathrm{SFR}}{2\pi R_{\mathrm{sf}}^2}.
\end{equation}
Here, $R_{\mathrm{sf}}$ is the effective size of all of the star-forming regions and clumps. Since nearly all the of a galaxy's star formation takes place in these regions, the numerator of the K-S relation is the global SFR of the galaxy. Substituting everything into Equation \ref{eq:alice_a_lambda}, we arrive at
\begin{equation}\label{eq:dust_model}
    A_\lambda \propto 10^{a\times\logoh}\times \left(\frac{\mathrm{SFR}}{R_{\mathrm{sf}}^2}\right)^{\left(\frac{1}{n}\right)}.
\end{equation}
Therefore, we find that the Balmer attenuation, \ABalmer, should be larger with increasing metallicity (higher dust-to-gas ratio), increasing SFR (more gas in the galaxy), and decreasing $R_{\mathrm{sf}}$ (more gas per unit area). Finally, this gives a prediction for the relation between metallicity and SFR that can control Balmer attenuation. At fixed mass, if metallicity decreases linearly with $\log_{10}\left(\mathrm{SFR}\right)$, we would find a constant Balmer attenuation. 

\begin{figure*}[tp]
\centering
\includegraphics[width=0.75\textwidth]{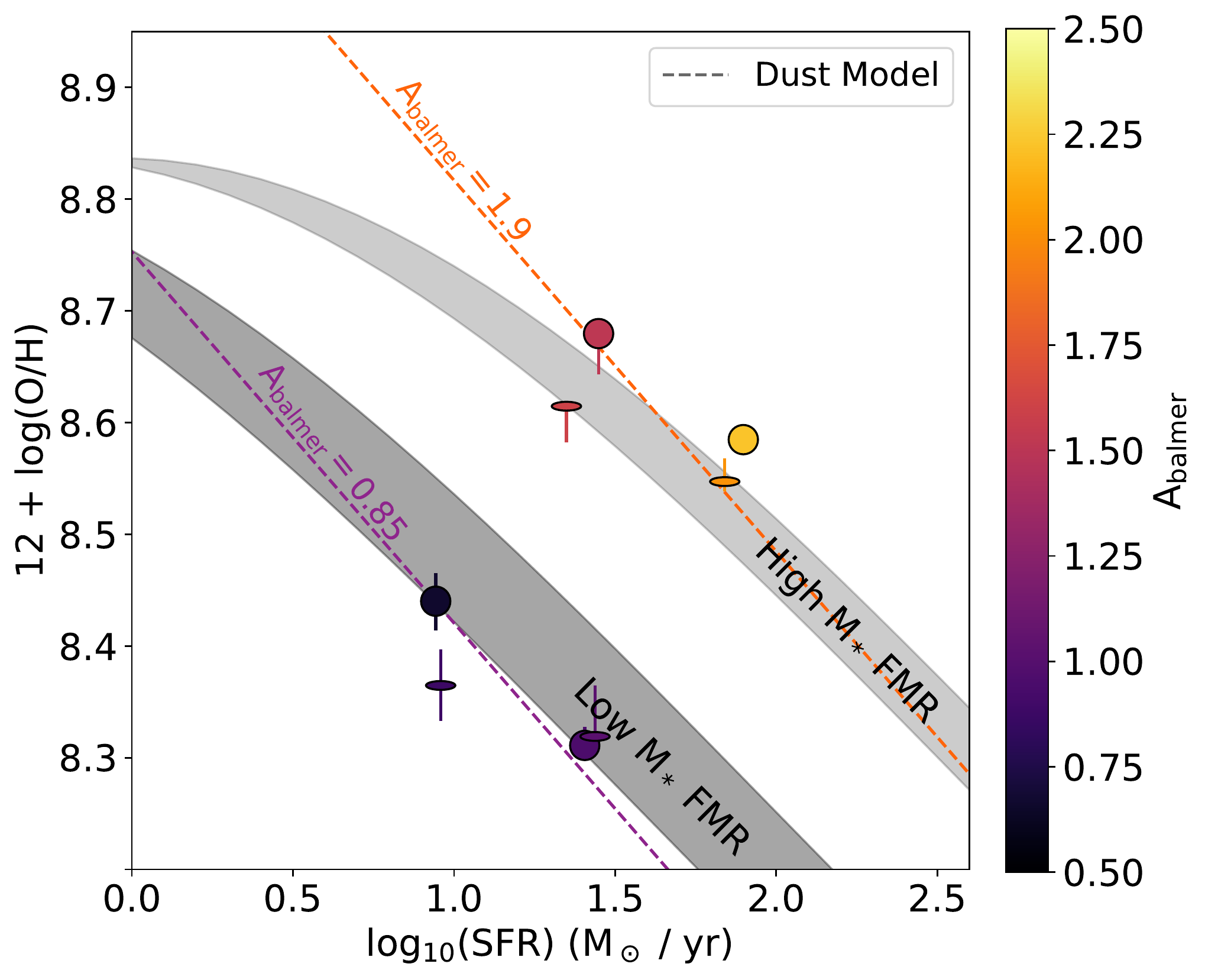}
\vglue -1pt
\caption{
Metallicity vs. SFR for the eight groups of galaxies, color-coded by stellar mass. Lines of constant Balmer attenuation generated from our dust model are shown (Equation \ref{eq:dust_model}, adopting $a=2.15, n=1.4$ and leaving $R_{\mathrm{sf}}$ free). Additionally, we display shaded regions of the FMR from \citet{sanders_mosdef_2021} that correspond to the mean masses of the low-mass and high-mass groups. The groups appear to fall near lines of constant attenuation despite their differing metallicities and SFRs. The slope of constant Balmer attenuation appears to be very similar to the slope of the FMR at fixed mass, so galaxies of similar mass will have similar Balmer attenuation regardless of their SFR or corresponding metallicity.}
\label{fig:metallicity_sfr}
\end{figure*}

Coincidentally, the FMR does indeed show that metallicity decreases roughly linearly with $\log_{10}\left(\mathrm{SFR}\right)$. To illustrate the relationship between the FMR and our dust model, we plot our model, along with our measured SFRs, metallicities, and the Fundamental Metallicity Relation (FMR), in Figure \ref{fig:metallicity_sfr}. We fix a constant of proportionality and solve for the best-fitting $R_\mathrm{sf}$ using least-squares for the low and high mass galaxies separately. The FMR is from \citet{sanders_mosdef_2021} for MOSDEF $z\approx 2.3$ and $z\approx 3.3$ star-forming galaxies. This relation is 
\begin{equation}
12 + \log_{10}\left(\mathrm{O/H}\right) = 8.80+0.188y-0.220y^2-0.0531y^3,
\end{equation}
where $y=\mu_{0.60}-10$ and $\mu_{0.60} = \log\left(\mathrm{\Mstar/\Msun}\right) - 0.6\times\log\left(\mathrm{SFR}/\mathrm{\Msun}\ \mathrm{ yr}^{-1}\right)$. 

Figure \ref{fig:metallicity_sfr} shows that the groups divide clearly by mass, each falling on their own roughly linear relation between metallicity and SFR as dictated by the FMR from \citet{sanders_mosdef_2021}. We also see lines of constant Balmer attenuation as predicted by our model in Equation \ref{eq:dust_model}. Most notably, we find that the slope of the FMR at fixed mass ($\approx -0.25$) is very similar to the slope of a constant Balmer attenuation according to the model ($-0.33$). Therefore, at fixed mass, any SFR has a corresponding metallicity from the FMR that results in a constant Balmer decrement. This correlation implies that galaxy mass is the only driver of Balmer attenuation, which is highlighted diagrammatically in Figure \ref{fig:dust_diagram_metallicity}.

Finally, we note that for either of the proposed dust geometries from Section \ref{subsec:implications_dust}, this dust model can still hold. Although the dust may be distributed among star-forming regions and clumps, the attenuating dust is roughly spherically symmetric. The assumptions for this model require only uniform distribution of dust and a spatially independent $\kappa_\lambda$, which are both plausible.

\section{Summary} \label{sec:summary}

In this paper, we use a sample of 308 star-forming galaxies at $1.37\leq z\leq 2.61$ from the MOSDEF survey to assess the relationship between dust properties and galaxy inclination. The galaxies have stellar masses of $9\leq\log\left(M_*\right)\leq11$ and measured axis ratios from deep HST/CANDELS imaging. We divide the data into eight groups such that, within each group, the galaxies have similar SFR, stellar mass, and axis ratio. In each group, we stack the spectra of all constituent galaxies. From these stacks, we measure emission lines to determine Balmer decrements and metallicities. We also compute median \AV and $\beta$ for each group. Below, we summarize our key findings:

\begin{itemize}
    \item All 308 galaxies have axis ratio measurements, with similar measurements in both F125W and F160W. Based on the distribution of these axis ratios, this sample of MOSDEF galaxies appears to be disky. 
    \item The Balmer decrement, as well as \AV and $\beta$, are independent of galaxy inclination. We also do not find trends in the UVJ distribution with inclination. Thus, face-on and edge-on galaxies have the same dust attenuation in our sample.
    \item The Balmer decrement, \AV, and $\beta$. all increase with increasing galaxy mass. Especially for the low-mass galaxies, Balmer decrement seems to be determined primarily by galaxy stellar mass, with no residual dependence on SFR or metallicity. The high mass galaxies may show a slight trend of increasing Balmer decrement with increasing SFR.
    \item The nebular attenuation is larger than the stellar attenuation, and this difference grows even larger at higher masses and SFRs. This finding implies that the nebular emission originates from more dusty regions than the stellar emission, on average, and the prevalence of these dustier regions increases with galaxy mass and SFR.
    \item The lack of variation in Balmer decrement, \AV, and $\beta$ with galaxy inclination supports a geometry in which the diffuse ISM does not play a large role in attenuation. We propose a three-component model consisting of very dusty and large star-forming clumps, typical star-forming regions, and minimal contributions from the diffuse ISM. In this model, dust attenuation does not show changes with inclination because the attenuation primarily occurs in (roughly spherical) star-forming regions and clumps. Due to low dust content in the ISM and high sSFR of the sample, we expect that both the nebular and stellar attenuation are independent of inclination. This model also predicts the trend that nebular and stellar attenuation increases with galaxy mass, since clumps are larger and more prevalent at higher galaxy masses. Finally, since more nebular emission originates from the clumps, and the clumps emit a larger fraction of the light at higher galaxy masses, the model also explains the growing differential attenuation between the nebular and stellar components as galaxy mass increases (see also \citet{reddy_mosdef_2020}).
    \item A fixed metal yield naturally explains why Balmer decrement solely depends on galaxy mass, with no residual dependence between Balmer decrement and SFR or metallicity. Balmer decrement only depends upon the dust column density, while SFR and metallicity relate to the gas content. Another way to understand this result is by examining the relationship between the gas content (related to SFR by the K-S relation) and dust-to-gas ratio (related to metallicity) of a galaxy. At fixed mass, a galaxy with higher SFR will have lower metallicity (i.e., the Fundamental Metallicity Relation), resulting in the same Balmer decrement as a galaxy with low SFR but higher metallicity. Equation \ref{eq:dust_model} describes this relationship quantitatively.

    % For example, a higher metallicity galaxy has lower SFR, and therefore less gas in total, so its higher dust-to-gas ratio leaves it with a similar amount of dust attenuation as a similar mass galaxy with high SFR and low metallicity.
    % Maybe add a bullet for the two-component
\end{itemize} 
The work above is based on stacked spectra. While stacking spectra is an incredibly powerful tool, it comes with some limitations. For example, rather than probing a continuous range of SFRs, we are only able to probe two bins --- information is washed out in the stacks. Rather than averaging over so many properties, we ideally would like to have accurate dust properties, SFR, mass, and axis ratio measurements for each individual galaxy. In particular, with these measurements, we could check the lowest sSFR galaxies to see if \AV and $\beta$ properties change with axis ratio, as we expect from our work. Similarly, examining the systems with the lowest axis ratios to search for increased dust attenuation would give more insight into the role of the diffuse ISM in dust attenuation. Neither of these tests are possible with current data, since the signal-to-noise of the emission features to measure Balmer decrements and metallicities are be too low. 

With higher quality data on each galaxy, we would no longer need to stack spectra, and therefore could obtain an even more accurate picture of the distribution of dust in star-forming galaxies during the epoch of peak star-formation in the universe. Fortunately, with the James Webb Space Telescope (JWST), it will finally be possible to obtain such spectroscopic samples and probe the dust distributions of high redshift galaxies in detail. Additionally, with the NIRSpec integral field unit on JWST, it will be possible to directly test our proposed three-component model. It can obtain Balmer decrements directly from the large star-forming clumps and from the regions around them to verify that the clumps are indeed responsible for stronger attenuation.

\begin{acknowledgments}
We thank the referee for a thorough and thoughtful report. BL thanks Daniel D. Kelson for the inspiration for this project. We would like to thank Robert Feldmann for useful discussions regarding the interpretation of results. This research is based on observations made with the NASA/ESA Hubble Space Telescope obtained from the Space Telescope Science Institute, which is operated by the Association of Universities for Research in Astronomy, Inc., under NASA contract NAS 5–26555. These observations are associated with program HST-AR-16141.001-A. This material is based upon work supported by the National Science Foundation Graduate Research Fellowship under Grant No. DGE 2146752. We also acknowledge support from NSF AAG grant Nos. AST- 1312780, 1312547, 1312764, 1313171, 2009313, and 2009085, grant No. AR-13907 from the Space Telescope Science Institute, and grant No. NNX16AF54G from the NASA ADAP program. We wish to recognize and acknowledge the very significant cultural role and reverence that the summit of Maunakea has always had within the indigenous Hawaiian community. We are most fortunate to have the opportunity to conduct observations from this mountain.
\end{acknowledgments}

%% IMPORTANT! The old "\acknowledgment" command has be depreciated. It was
%% not robust enough to handle our new dual anonymous review requirements and
%% thus been replaced with the acknowledgment environment. If you try to 
%% compile with \acknowledgment you will get an error print to the screen
%% and in the compiled pdf.

%% To help institutions obtain information on the effectiveness of their 
%% telescopes the AAS Journals has created a group of keywords for telescope 
%% facilities.
%
%% Following the acknowledgments section, use the following syntax and the
%% \facility{} or \facilities{} macros to list the keywords of facilities used 
%% in the research for the paper.  Each keyword is check against the master 
%% list during copy editing.  Individual instruments can be provided in 
%% parentheses, after the keyword, but they are not verified.

\vspace{5mm}
\facilities{HST(STIS), Keck}

%% Similar to \facility{}, there is the optional \software command to allow 
%% authors a place to specify which programs were used during the creation of 
%% the manuscript. Authors should list each code and include either a
%% citation or url to the code inside ()s when available.

% \software{astropy \citep{2013A&A...558A..33A,2018AJ....156..123A},  
%           Cloudy \citep{2013RMxAA..49..137F}, 
%           Source Extractor \citep{1996A&AS..117..393B}
%           }

%% Appendix material should be preceded with a single \appendix command.
%% There should be a \section command for each appendix. Mark appendix
%% subsections with the same markup you use in the main body of the paper.

%% Each Appendix (indicated with \section) will be lettered A, B, C, etc.
%% The equation counter will reset when it encounters the \appendix
%% command and will number appendix equations (A1), (A2), etc. The
%% Figure and Table counter will not reset.

%% For this sample we use BibTeX plus aasjournals.bst to generate the
%% the bibliography. The sample631.bib file was populated from ADS. To
%% get the citations to show in the compiled file do the following:
%%
%% pdflatex sample631.tex
%% bibtext sample631
%% pdflatex sample631.tex
%% pdflatex sample631.tex

\typeout{}\bibliography{MOSDEF}{}

\begin{thebibliography}{}
\expandafter\ifx\csname natexlab\endcsname\relax\def\natexlab#1{#1}\fi
\providecommand{\url}[1]{\href{#1}{#1}}
\providecommand{\dodoi}[1]{doi:~\href{http://doi.org/#1}{\nolinkurl{#1}}}
\providecommand{\doeprint}[1]{\href{http://ascl.net/#1}{\nolinkurl{http://ascl.net/#1}}}
\providecommand{\doarXiv}[1]{\href{https://arxiv.org/abs/#1}{\nolinkurl{https://arxiv.org/abs/#1}}}

\bibitem[{Azadi {et~al.}(2017)Azadi, Coil, Aird, Reddy, Shapley, Freeman,
  Kriek, Leung, Mobasher, Price, Sanders, Shivaei, \&
  Siana}]{azadi_mosdef_2017}
Azadi, M., Coil, A.~L., Aird, J., {et~al.} 2017, The Astrophysical Journal,
  835, 27, \dodoi{10.3847/1538-4357/835/1/27}

\bibitem[{Barro {et~al.}(2016)Barro, Kriek, Pérez-González, Trump, Koo,
  Faber, Dekel, Primack, Guo, Kocevski, Muñoz-Mateos, Rujopakarn, \&
  Seth}]{barro_sub-kiloparsec_2016}
Barro, G., Kriek, M., Pérez-González, P.~G., {et~al.} 2016, The Astrophysical
  Journal, 827, L32, \dodoi{10.3847/2041-8205/827/2/L32}

\bibitem[{Battisti {et~al.}(2017)Battisti, Calzetti, \&
  Chary}]{battisti_characterizing_2017}
Battisti, A.~J., Calzetti, D., \& Chary, R.-R. 2017, The Astrophysical Journal,
  851, 90, \dodoi{10.3847/1538-4357/aa9a43}

\bibitem[{Bian {et~al.}(2018)Bian, Kewley, \& Dopita}]{bian_direct_2018}
Bian, F., Kewley, L.~J., \& Dopita, M.~A. 2018, The Astrophysical Journal, 859,
  175, \dodoi{10.3847/1538-4357/aabd74}

\bibitem[{Brammer {et~al.}(2009)Brammer, Whitaker, van Dokkum, Marchesini,
  Labbé, Franx, Kriek, Quadri, Illingworth, Lee, Muzzin, \&
  Rudnick}]{brammer_dead_2009}
Brammer, G.~B., Whitaker, K.~E., van Dokkum, P.~G., {et~al.} 2009, The
  Astrophysical Journal, 706, L173, \dodoi{10.1088/0004-637X/706/1/L173}

\bibitem[{Brammer {et~al.}(2012)Brammer, van Dokkum, Franx, Fumagalli, Patel,
  Rix, Skelton, Kriek, Nelson, Schmidt, Bezanson, da~Cunha, Erb, Fan,
  Förster~Schreiber, Illingworth, Labbé, Leja, Lundgren, Magee, Marchesini,
  McCarthy, Momcheva, Muzzin, Quadri, Steidel, Tal, Wake, Whitaker, \&
  Williams}]{brammer_3d-hst_2012}
Brammer, G.~B., van Dokkum, P.~G., Franx, M., {et~al.} 2012, The Astrophysical
  Journal Supplement Series, 200, 13, \dodoi{10.1088/0067-0049/200/2/13}

\bibitem[{Calzetti {et~al.}(2000)Calzetti, Armus, Bohlin, Kinney, Koornneef, \&
  Storchi‐Bergmann}]{calzetti_dust_2000}
Calzetti, D., Armus, L., Bohlin, R.~C., {et~al.} 2000, The Astrophysical
  Journal, 533, 682, \dodoi{10.1086/308692}

\bibitem[{Cardelli {et~al.}(1989)Cardelli, Clayton, \&
  Mathis}]{cardelli_relationship_1989}
Cardelli, J.~A., Clayton, G.~C., \& Mathis, J.~S. 1989, The Astrophysical
  Journal, 345, 245, \dodoi{10.1086/167900}

\bibitem[{Chabrier(2003)}]{chabrier_galactic_2003}
Chabrier, G. 2003, Publications of the Astronomical Society of the Pacific,
  115, 763, \dodoi{10.1086/376392}

\bibitem[{Charlot \& Fall(2000)}]{charlot_simple_2000}
Charlot, S., \& Fall, S.~M. 2000, The Astrophysical Journal, 539, 718,
  \dodoi{10.1086/309250}

\bibitem[{Coil {et~al.}(2015)Coil, Aird, Reddy, Shapley, Kriek, Siana,
  Mobasher, Freeman, Price, \& Shivaei}]{coil_mosdef_2015}
Coil, A.~L., Aird, J., Reddy, N., {et~al.} 2015, The Astrophysical Journal,
  801, 35, \dodoi{10.1088/0004-637X/801/1/35}

\bibitem[{Conroy \& Gunn(2010)}]{conroy_propagation_2010}
Conroy, C., \& Gunn, J.~E. 2010, The Astrophysical Journal, 712, 833,
  \dodoi{10.1088/0004-637X/712/2/833}

\bibitem[{Conroy {et~al.}(2009)Conroy, Gunn, \&
  White}]{conroy_propagation_2009}
Conroy, C., Gunn, J.~E., \& White, M. 2009, The Astrophysical Journal, 699,
  486, \dodoi{10.1088/0004-637X/699/1/486}

\bibitem[{De~Vis {et~al.}(2019)De~Vis, Jones, Viaene, Casasola, Clark, Baes,
  Bianchi, Cassara, Davies, De~Looze, Galametz, Galliano, Lianou, Madden,
  Manilla-Robles, Mosenkov, Nersesian, Roychowdhury, Xilouris, \&
  Ysard}]{de_vis_systematic_2019}
De~Vis, P., Jones, A., Viaene, S., {et~al.} 2019, Astronomy \&amp;
  Astrophysics, Volume 623, id.A5, 25 pp., 623, A5,
  \dodoi{10.1051/0004-6361/201834444}

\bibitem[{Elmegreen \& Elmegreen(2006)}]{elmegreen_observations_2006}
Elmegreen, B.~G., \& Elmegreen, D.~M. 2006, The Astrophysical Journal, 650,
  644, \dodoi{10.1086/507578}

\bibitem[{Fetherolf {et~al.}(2023)Fetherolf, Reddy, Shapley, Kriek, Siana,
  Coil, Mobasher, Freeman, Price, Sanders, Shivaei, Azadi, de~Groot, Leung, \&
  Zick}]{fetherolf_mosdef_2023}
Fetherolf, T., Reddy, N.~A., Shapley, A.~E., {et~al.} 2023, Monthly Notices of
  the Royal Astronomical Society, 518, 4214, \dodoi{10.1093/mnras/stac3362}

\bibitem[{Förster~Schreiber {et~al.}(2011)Förster~Schreiber, Shapley, Genzel,
  Bouché, Cresci, Davies, Erb, Genel, Lutz, Newman, Shapiro, Steidel,
  Sternberg, \& Tacconi}]{schreiber_constraints_2011}
Förster~Schreiber, N.~M., Shapley, A.~E., Genzel, R., {et~al.} 2011, The
  Astrophysical Journal, 739, 45, \dodoi{10.1088/0004-637X/739/1/45}

\bibitem[{Garn \& Best(2010)}]{garn_predicting_2010}
Garn, T., \& Best, P.~N. 2010, Monthly Notices of the Royal Astronomical
  Society, 409, 421, \dodoi{10.1111/j.1365-2966.2010.17321.x}

\bibitem[{Genzel {et~al.}(2008)Genzel, Burkert, Bouché, Cresci,
  Förster~Schreiber, Shapley, Shapiro, Tacconi, Buschkamp, Cimatti, Daddi,
  Davies, Eisenhauer, Erb, Genel, Gerhard, Hicks, Lutz, Naab, Ott, Rabien,
  Renzini, Steidel, Sternberg, \& Lilly}]{genzel_rings_2008}
Genzel, R., Burkert, A., Bouché, N., {et~al.} 2008, The Astrophysical Journal,
  687, 59, \dodoi{10.1086/591840}

\bibitem[{Grogin {et~al.}(2011)Grogin, Kocevski, Faber, Ferguson, Koekemoer,
  Riess, Acquaviva, Alexander, Almaini, Ashby, Barden, Bell, Bournaud, Brown,
  Caputi, Casertano, Cassata, Castellano, Challis, Chary, Cheung, Cirasuolo,
  Conselice, Cooray, Croton, Daddi, Dahlen, Davé, de~Mello, Dekel, Dickinson,
  Dolch, Donley, Dunlop, Dutton, Elbaz, Fazio, Filippenko, Finkelstein,
  Fontana, Gardner, Garnavich, Gawiser, Giavalisco, Grazian, Guo, Hathi,
  Häussler, Hopkins, Huang, Huang, Jha, Kartaltepe, Kirshner, Koo, Lai, Lee,
  Li, Lotz, Lucas, Madau, McCarthy, McGrath, McIntosh, McLure, Mobasher,
  Moustakas, Mozena, Nandra, Newman, Niemi, Noeske, Papovich, Pentericci, Pope,
  Primack, Rajan, Ravindranath, Reddy, Renzini, Rix, Robaina, Rodney, Rosario,
  Rosati, Salimbeni, Scarlata, Siana, Simard, Smidt, Somerville, Spinrad,
  Straughn, Strolger, Telford, Teplitz, Trump, van~der Wel, Villforth,
  Wechsler, Weiner, Wiklind, Wild, Wilson, Wuyts, Yan, \&
  Yun}]{grogin_candels_2011}
Grogin, N.~A., Kocevski, D.~D., Faber, S.~M., {et~al.} 2011, The Astrophysical
  Journal Supplement Series, 197, 35, \dodoi{10.1088/0067-0049/197/2/35}

\bibitem[{Hao {et~al.}(2011)Hao, Kennicutt, Johnson, Calzetti, Dale, \&
  Moustakas}]{hao_dust-corrected_2011}
Hao, C.-N., Kennicutt, R.~C., Johnson, B.~D., {et~al.} 2011, The Astrophysical
  Journal, 741, 124, \dodoi{10.1088/0004-637X/741/2/124}

\bibitem[{Kennicutt(1998)}]{kennicutt_global_1998}
Kennicutt, Jr., R.~C. 1998, The Astrophysical Journal, 498, 541,
  \dodoi{10.1086/305588}

\bibitem[{Koekemoer {et~al.}(2011)Koekemoer, Faber, Ferguson, Grogin, Kocevski,
  Koo, Lai, Lotz, Lucas, McGrath, Ogaz, Rajan, Riess, Rodney, Strolger,
  Casertano, Castellano, Dahlen, Dickinson, Dolch, Fontana, Giavalisco,
  Grazian, Guo, Hathi, Huang, van~der Wel, Yan, Acquaviva, Almaini, Ashby,
  Barden, Bell, Bournaud, Brown, Caputi, Cassata, Challis, Chary, Cheung,
  Cirasuolo, Conselice, Cooray, Croton, Daddi, Davé, de~Mello, de~Ravel,
  Dekel, Donley, Dunlop, Dutton, Elbaz, Fazio, Filippenko, Finkelstein, Frazer,
  Gardner, Garnavich, Gawiser, Gruetzbauch, Hartley, Häussler, Herrington,
  Hopkins, Huang, Jha, Johnson, Kartaltepe, Khostovan, Kirshner, Lani, Lee, Li,
  Madau, McCarthy, McIntosh, McLure, McPartland, Mobasher, Moreira, Mortlock,
  Moustakas, Mozena, Nandra, Newman, Nielsen, Niemi, Noeske, Papovich,
  Pentericci, Pope, Primack, Ravindranath, Reddy, Renzini, Rix, Robaina,
  Rosario, Rosati, Salimbeni, Scarlata, Siana, Simard, Smidt, Snyder,
  Somerville, Spinrad, Straughn, Telford, Teplitz, Trump, Vargas, Villforth,
  Wagner, Wandro, Wechsler, Weiner, Wiklind, Wild, Wilson, Wuyts, \&
  Yun}]{koekemoer_candels_2011}
Koekemoer, A.~M., Faber, S.~M., Ferguson, H.~C., {et~al.} 2011, The
  Astrophysical Journal Supplement Series, 197, 36,
  \dodoi{10.1088/0067-0049/197/2/36}

\bibitem[{Kriek {et~al.}(2009)Kriek, van Dokkum, Labbé, Franx, Illingworth,
  Marchesini, \& Quadri}]{kriek_ultra-deep_2009}
Kriek, M., van Dokkum, P.~G., Labbé, I., {et~al.} 2009, The Astrophysical
  Journal, 700, 221, \dodoi{10.1088/0004-637X/700/1/221}

\bibitem[{Kriek {et~al.}(2015)Kriek, Shapley, Reddy, Siana, Coil, Mobasher,
  Freeman, de~Groot, Price, Sanders, Shivaei, Brammer, Momcheva, Skelton, van
  Dokkum, Whitaker, Aird, Azadi, Kassis, Bullock, Conroy, Dave, Keres, \&
  Krumholz}]{kriek_mosfire_2015}
Kriek, M., Shapley, A.~E., Reddy, N.~A., {et~al.} 2015, The Astrophysical
  Journal Supplement Series, 218, 15, \dodoi{10.1088/0067-0049/218/2/15}

\bibitem[{Labbé {et~al.}(2005)Labbé, Huang, Franx, Rudnick, Barmby, Daddi,
  van Dokkum, Fazio, Förster~Schreiber, Moorwood, Rix, Röttgering, Trujillo,
  \& van~der Werf}]{labbe_irac_2005}
Labbé, I., Huang, J., Franx, M., {et~al.} 2005, The Astrophysical Journal,
  624, L81, \dodoi{10.1086/430700}

\bibitem[{Lambas {et~al.}(1992)Lambas, Maddox, \& Loveday}]{lambas_true_1992}
Lambas, D.~G., Maddox, S.~J., \& Loveday, J. 1992, Monthly Notices of the Royal
  Astronomical Society, 258, 404, \dodoi{10.1093/mnras/258.2.404}

\bibitem[{Liu {et~al.}(2008)Liu, Shapley, Coil, Brinchmann, \&
  Ma}]{liu_metallicities_2008}
Liu, X., Shapley, A.~E., Coil, A.~L., Brinchmann, J., \& Ma, C.-P. 2008, The
  Astrophysical Journal, 678, 758, \dodoi{10.1086/529030}

\bibitem[{Mannucci {et~al.}(2010)Mannucci, Cresci, Maiolino, Marconi, \&
  Gnerucci}]{mannucci_fundamental_2010}
Mannucci, F., Cresci, G., Maiolino, R., Marconi, A., \& Gnerucci, A. 2010,
  Monthly Notices of the Royal Astronomical Society, 408, 2115,
  \dodoi{10.1111/j.1365-2966.2010.17291.x}

\bibitem[{McLean {et~al.}(2012)McLean, Steidel, Epps, Konidaris, Matthews,
  Adkins, Aliado, Brims, Canfield, Cromer, Fucik, Kulas, Mace, Magnone,
  Rodriguez, Rudie, Trainor, Wang, Weber, \& Weiss}]{mclean_mosfire_2012}
McLean, I.~S., Steidel, C.~C., Epps, H.~W., {et~al.} 2012, 8446, 84460J,
  \dodoi{10.1117/12.924794}

\bibitem[{Misiriotis {et~al.}(2001)Misiriotis, Popescu, Tuffs, \&
  Kylafis}]{misiriotis_modeling_2001}
Misiriotis, A., Popescu, C.~C., Tuffs, R., \& Kylafis, N.~D. 2001, Astronomy
  and Astrophysics, v.372, p.775-783 (2001), 372, 775,
  \dodoi{10.1051/0004-6361:20010568}

\bibitem[{Momcheva {et~al.}(2016)Momcheva, Brammer, van Dokkum, Skelton,
  Whitaker, Nelson, Fumagalli, Maseda, Leja, Franx, Rix, Bezanson, Da~Cunha,
  Dickey, Schreiber, Illingworth, Kriek, Labbé, Lange, Lundgren, Magee,
  Marchesini, Oesch, Pacifici, Patel, Price, Tal, Wake, van~der Wel, \&
  Wuyts}]{momcheva_3d-hst_2016}
Momcheva, I.~G., Brammer, G.~B., van Dokkum, P.~G., {et~al.} 2016, The
  Astrophysical Journal Supplement Series, 225, 27,
  \dodoi{10.3847/0067-0049/225/2/27}

\bibitem[{Patel {et~al.}(2012)Patel, Holden, Kelson, Franx, van~der Wel, \&
  Illingworth}]{patel_uvj_2012}
Patel, S.~G., Holden, B.~P., Kelson, D.~D., {et~al.} 2012, The Astrophysical
  Journal, 748, L27, \dodoi{10.1088/2041-8205/748/2/L27}

\bibitem[{Peng {et~al.}(2010)Peng, Ho, Impey, \& Rix}]{peng_detailed_2010}
Peng, C.~Y., Ho, L.~C., Impey, C.~D., \& Rix, H.-W. 2010, The Astronomical
  Journal, 139, 2097, \dodoi{10.1088/0004-6256/139/6/2097}

\bibitem[{Popping {et~al.}(2022)Popping, Shivaei, Sanders, Jones, Pope, Reddy,
  Shapley, Coil, \& Kriek}]{popping_dust--gas_2022}
Popping, G., Shivaei, I., Sanders, R.~L., {et~al.} 2022, arXiv:2204.08483
  [astro-ph].
\newblock \url{http://arxiv.org/abs/2204.08483}

\bibitem[{Price {et~al.}(2014)Price, Kriek, Brammer, Conroy, Schreiber, Franx,
  Fumagalli, Lundgren, Momcheva, Nelson, Skelton, van Dokkum, Whitaker, \&
  Wuyts}]{price_direct_2014}
Price, S.~H., Kriek, M., Brammer, G.~B., {et~al.} 2014, The Astrophysical
  Journal, 788, 86, \dodoi{10.1088/0004-637X/788/1/86}

\bibitem[{Price {et~al.}(2016)Price, Kriek, Shapley, Reddy, Freeman, Coil,
  de~Groot, Shivaei, Siana, Azadi, Barro, Mobasher, Sanders, \&
  Zick}]{price_mosdef_2016}
Price, S.~H., Kriek, M., Shapley, A.~E., {et~al.} 2016, The Astrophysical
  Journal, 819, 80, \dodoi{10.3847/0004-637X/819/1/80}

\bibitem[{Price {et~al.}(2020)Price, Kriek, Barro, Shapley, Reddy, Freeman,
  Coil, Shivaei, Azadi, de~Groot, Siana, Mobasher, Sanders, Leung, Fetherolf,
  Zick, Übler, \& Schreiber}]{price_mosdef_2020}
Price, S.~H., Kriek, M., Barro, G., {et~al.} 2020, The Astrophysical Journal,
  894, 91, \dodoi{10.3847/1538-4357/ab7990}

\bibitem[{Reddy {et~al.}(2015)Reddy, Kriek, Shapley, Freeman, Siana, Coil,
  Mobasher, Price, Sanders, \& Shivaei}]{reddy_mosdef_2015}
Reddy, N.~A., Kriek, M., Shapley, A.~E., {et~al.} 2015, The Astrophysical
  Journal, 806, 259, \dodoi{10.1088/0004-637X/806/2/259}

\bibitem[{Reddy {et~al.}(2020)Reddy, Shapley, Kriek, Steidel, Shivaei, Sanders,
  Mobasher, Coil, Siana, Freeman, Azadi, Fetherolf, Leung, Price, \&
  Zick}]{reddy_mosdef_2020}
Reddy, N.~A., Shapley, A.~E., Kriek, M., {et~al.} 2020, The Astrophysical
  Journal, 902, 123, \dodoi{10.3847/1538-4357/abb674}

\bibitem[{Rodríguez \& Padilla(2013)}]{rodriguez_intrinsic_2013}
Rodríguez, S., \& Padilla, N.~D. 2013, Monthly Notices of the Royal
  Astronomical Society, 434, 2153, \dodoi{10.1093/mnras/stt1168}

\bibitem[{Salim \& Narayanan(2020)}]{salim_dust_2020}
Salim, S., \& Narayanan, D. 2020, arXiv:2001.03181 [astro-ph].
\newblock \url{http://arxiv.org/abs/2001.03181}

\bibitem[{Sanders {et~al.}(2015)Sanders, Shapley, Kriek, Reddy, Freeman, Coil,
  Siana, Mobasher, Shivaei, Price, \& de~Groot}]{sanders_mosdef_2015}
Sanders, R.~L., Shapley, A.~E., Kriek, M., {et~al.} 2015, The Astrophysical
  Journal, 799, 138, \dodoi{10.1088/0004-637X/799/2/138}

\bibitem[{Sanders {et~al.}(2018)Sanders, Shapley, Kriek, Freeman, Reddy, Siana,
  Coil, Mobasher, Davé, Shivaei, Azadi, Price, Leung, Fetherolf, de~Groot,
  Zick, Fornasini, \& Barro}]{sanders_mosdef_2018}
---. 2018, The Astrophysical Journal, 858, 99, \dodoi{10.3847/1538-4357/aabcbd}

\bibitem[{Sanders {et~al.}(2020)Sanders, Shapley, Reddy, Kriek, Siana, Coil,
  Mobasher, Shivaei, Freeman, Azadi, Price, Leung, Fetherolf, de~Groot, Zick,
  Fornasini, \& Barro}]{sanders_mosdef_2020}
Sanders, R.~L., Shapley, A.~E., Reddy, N.~A., {et~al.} 2020, Monthly Notices of
  the Royal Astronomical Society, 491, 1427, \dodoi{10.1093/mnras/stz3032}

\bibitem[{Sanders {et~al.}(2021)Sanders, Shapley, Jones, Reddy, Kriek, Siana,
  Coil, Mobasher, Shivaei, Davé, Azadi, Price, Leung, Freeman, Fetherolf,
  de~Groot, Zick, \& Barro}]{sanders_mosdef_2021}
Sanders, R.~L., Shapley, A.~E., Jones, T., {et~al.} 2021, The Astrophysical
  Journal, 914, 19, \dodoi{10.3847/1538-4357/abf4c1}

\bibitem[{Shapley {et~al.}(2020)Shapley, Cullen, Dunlop, McLure, Kriek, Reddy,
  \& Sanders}]{shapley_first_2020}
Shapley, A.~E., Cullen, F., Dunlop, J.~S., {et~al.} 2020, The Astrophysical
  Journal, 903, L16, \dodoi{10.3847/2041-8213/abc006}

\bibitem[{Shapley {et~al.}(2022)Shapley, Sanders, Salim, Reddy, Kriek,
  Mobasher, Coil, Siana, Price, Shivaei, Dunlop, McLure, \&
  Cullen}]{shapley_mosfire_2022}
Shapley, A.~E., Sanders, R.~L., Salim, S., {et~al.} 2022, The Astrophysical
  Journal, 926, 145, \dodoi{10.3847/1538-4357/ac4742}

\bibitem[{Shivaei {et~al.}(2015)Shivaei, Reddy, Shapley, Kriek, Siana,
  Mobasher, Coil, Freeman, Sanders, Price, de~Groot, \&
  Azadi}]{shivaei_mosdef_2015}
Shivaei, I., Reddy, N.~A., Shapley, A.~E., {et~al.} 2015, The Astrophysical
  Journal, 815, 98, \dodoi{10.1088/0004-637X/815/2/98}

\bibitem[{Shivaei {et~al.}(2016)Shivaei, Kriek, Reddy, Shapley, Barro, Conroy,
  Coil, Freeman, Mobasher, Siana, Sanders, Price, Azadi, Pasha, \&
  Inami}]{shivaei_mosdef_2016}
Shivaei, I., Kriek, M., Reddy, N.~A., {et~al.} 2016, The Astrophysical Journal,
  820, L23, \dodoi{10.3847/2041-8205/820/2/L23}

\bibitem[{Simons {et~al.}(2017)Simons, Kassin, Weiner, Faber, Trump, Heckman,
  Koo, Pacifici, Primack, Snyder, \& de~la Vega}]{simons_z2_2017}
Simons, R.~C., Kassin, S.~A., Weiner, B.~J., {et~al.} 2017, The Astrophysical
  Journal, 843, 46, \dodoi{10.3847/1538-4357/aa740c}

\bibitem[{Skelton {et~al.}(2014)Skelton, Whitaker, Momcheva, Brammer, van
  Dokkum, Labbe, Franx, van~der Wel, Bezanson, Da~Cunha, Fumagalli, Schreiber,
  Kriek, Leja, Lundgren, Magee, Marchesini, Maseda, Nelson, Oesch, Pacifici,
  Patel, Price, Rix, Tal, Wake, \& Wuyts}]{skelton_3d-hst_2014}
Skelton, R.~E., Whitaker, K.~E., Momcheva, I.~G., {et~al.} 2014, The
  Astrophysical Journal Supplement Series, 214, 24,
  \dodoi{10.1088/0067-0049/214/2/24}

\bibitem[{Steidel {et~al.}(2014)Steidel, Rudie, Strom, Pettini, Reddy, Shapley,
  Trainor, Erb, Turner, Konidaris, Kulas, Mace, Matthews, \&
  McLean}]{steidel_strong_2014}
Steidel, C.~C., Rudie, G.~C., Strom, A.~L., {et~al.} 2014, The Astrophysical
  Journal, 795, 165, \dodoi{10.1088/0004-637X/795/2/165}

\bibitem[{Swinbank {et~al.}(2012)Swinbank, Smail, Sobral, Theuns, Best, \&
  Geach}]{swinbank_properties_2012}
Swinbank, A.~M., Smail, I., Sobral, D., {et~al.} 2012, The Astrophysical
  Journal, 760, 130, \dodoi{10.1088/0004-637X/760/2/130}

\bibitem[{Tadaki {et~al.}(2014)Tadaki, Kodama, Tanaka, Hayashi, Koyama, \&
  Shimakawa}]{tadaki_nature_2014}
Tadaki, K.-i., Kodama, T., Tanaka, I., {et~al.} 2014, The Astrophysical
  Journal, 780, 77, \dodoi{10.1088/0004-637X/780/1/77}

\bibitem[{Theios {et~al.}(2019)Theios, Steidel, Strom, Rudie, Trainor, \&
  Reddy}]{theios_dust_2019}
Theios, R.~L., Steidel, C.~C., Strom, A.~L., {et~al.} 2019, The Astrophysical
  Journal, 871, 128, \dodoi{10.3847/1538-4357/aaf386}

\bibitem[{Tremonti {et~al.}(2004)Tremonti, Heckman, Kauffmann, Brinchmann,
  Charlot, White, Seibert, Peng, Schlegel, Uomoto, Fukugita, \&
  Brinkmann}]{tremonti_origin_2004}
Tremonti, C.~A., Heckman, T.~M., Kauffmann, G., {et~al.} 2004, The
  Astrophysical Journal, 613, 898, \dodoi{10.1086/423264}

\bibitem[{van~der Wel {et~al.}(2012)van~der Wel, Bell, Häussler, McGrath,
  Chang, Guo, McIntosh, Rix, Barden, Cheung, Faber, Ferguson, Galametz, Grogin,
  Hartley, Kartaltepe, Kocevski, Koekemoer, Lotz, Mozena, Peth, \&
  Peng}]{van_der_wel_structural_2012}
van~der Wel, A., Bell, E.~F., Häussler, B., {et~al.} 2012, The Astrophysical
  Journal Supplement Series, 203, 24, \dodoi{10.1088/0067-0049/203/2/24}

\bibitem[{Whitaker {et~al.}(2014)Whitaker, Franx, Leja, van Dokkum, Henry,
  Skelton, Fumagalli, Momcheva, Brammer, Labbé, Nelson, \&
  Rigby}]{whitaker_constraining_2014}
Whitaker, K.~E., Franx, M., Leja, J., {et~al.} 2014, The Astrophysical Journal,
  795, 104, \dodoi{10.1088/0004-637X/795/2/104}

\bibitem[{Wild {et~al.}(2011)Wild, Charlot, Brinchmann, Heckman, Vince,
  Pacifici, \& Chevallard}]{wild_empirical_2011}
Wild, V., Charlot, S., Brinchmann, J., {et~al.} 2011, Monthly Notices of the
  Royal Astronomical Society, 417, 1760,
  \dodoi{10.1111/j.1365-2966.2011.19367.x}

\bibitem[{Williams {et~al.}(2009)Williams, Quadri, Franx, van Dokkum, \&
  Labbé}]{williams_detection_2009}
Williams, R.~J., Quadri, R.~F., Franx, M., van Dokkum, P., \& Labbé, I. 2009,
  The Astrophysical Journal, 691, 1879, \dodoi{10.1088/0004-637X/691/2/1879}

\bibitem[{Wuyts {et~al.}(2007)Wuyts, Labbé, Franx, Rudnick, van Dokkum, Fazio,
  Förster~Schreiber, Huang, Moorwood, Rix, Röttgering, \& van~der
  Werf}]{wuyts_what_2007}
Wuyts, S., Labbé, I., Franx, M., {et~al.} 2007, The Astrophysical Journal,
  655, 51, \dodoi{10.1086/509708}

\bibitem[{Wuyts {et~al.}(2012)Wuyts, Schreiber, Genzel, Guo, Barro, Bell,
  Dekel, Faber, Ferguson, Giavalisco, Grogin, Hathi, Huang, Kocevski,
  Koekemoer, Koo, Lotz, Lutz, McGrath, Newman, Rosario, Saintonge, Tacconi,
  Weiner, \& van~der Wel}]{wuyts_smoother_2012}
Wuyts, S., Schreiber, N. M.~F., Genzel, R., {et~al.} 2012, The Astrophysical
  Journal, 753, 114, \dodoi{10.1088/0004-637X/753/2/114}

\bibitem[{Yip {et~al.}(2010)Yip, Szalay, Wyse, Dobos, Budavári, \&
  Csabai}]{yip_extinction_2010}
Yip, C.-W., Szalay, A.~S., Wyse, R. F.~G., {et~al.} 2010, The Astrophysical
  Journal, 709, 780, \dodoi{10.1088/0004-637X/709/2/780}

\bibitem[{Yuan {et~al.}(2021)Yuan, Lu, Shen, \& Boquien}]{yuan_asymmetry_2021}
Yuan, F.-T., Lu, J., Shen, S., \& Boquien, M. 2021, The Astrophysical Journal,
  911, 145, \dodoi{10.3847/1538-4357/abec76}

\bibitem[{Zuckerman {et~al.}(2021)Zuckerman, Belli, Leja, \&
  Tacchella}]{zuckerman_reproducing_2021}
Zuckerman, L.~D., Belli, S., Leja, J., \& Tacchella, S. 2021, The Astrophysical
  Journal, 922, L32, \dodoi{10.3847/2041-8213/ac3831}

\end{thebibliography}
\bibliographystyle{aasjournal}

%% This command is needed to show the entire author+affiliation list when
%% the collaboration and author truncation commands are used.  It has to
%% go at the end of the manuscript.
%\allauthors

%% Include this line if you are using the \added, \replaced, \deleted
%% commands to see a summary list of all changes at the end of the article.
%\listofchanges

\end{document}